\documentstyle [epsf,epsfig,multicol,aps,pre] {revtex}
\newcommand{\p}[2]    {p_{#1}^{(#2)}}
\newcommand{\ct}[1]   {c_t^{(#1)}}
\newcommand{\pgti}[1]    {\tilde p_g^{(#1)}}
\newcommand{\psti}[1]    {\tilde p_s^{(#1)}}
\newcommand{\ctti}[1]   {\tilde c_t^{(#1)}}
\newcommand{\cd}[1]   {c_d^{(#1)}}{\rm }
\newcommand{\fcat}   {f_{cat}^{(0)}}
\begin{document}
\draft
%

% ******************************************************************************
%
   \title  
 {Modeling  oscillatory Microtubule--Polymerization}

%
% ******************************************************************************

%
   \author  {Martin Hammele and Walter Zimmermann}
%
% ******************************************************************************

%
   \address { Theoretical Physics, University of the Saarland, 
D-66041 Saarbr{\"u}cken, Germany \\
%FORUM Modellierung and Institut f{\"u}r Festk{\"o}rperforschung,
%             Forschungszentrum J{\"u}lich,
%             D-52425 J{\"u}lich,
%              Germany 
 }
%
% ******************************************************************************

%

   \date  {\today}
%

% *****************************************************************************
%

\maketitle
%

% ******************************************************************************

%   Abstract

% ******************************************************************************

%

\begin{abstract}
Polymerization of microtubules is  ubiquitous in  biological cells
and under certain conditions it 
becomes  oscillatory in time.
Here simple reaction models 
are analyzed that capture  such oscillations
as well as the length distribution of microtubules.
 We assume
reaction conditions that are stationary
over many oscillation periods, and it is a  
Hopf bifurcation that leads to a persistent 
oscillatory microtubule polymerization in these models.
Analytical expressions  are derived for the threshold of  the bifurcation 
and the oscillation frequency  in terms of
reaction rates as well as typical trends of their parameter dependence 
are presented. Both, a catastrophe rate that depends on the 
density of {\it guanosine triphosphate} (GTP) liganded tubulin dimers
and  a   delay reaction, such as 
the depolymerization of shrinking microtubules
or the decay of oligomers,  support oscillations.
For a tubulin dimer concentration below the threshold 
oscillatory microtubule  polymerization occurs transiently  on the route
to a stationary state, as shown by numerical
solutions of the model equations. 
Close to threshold a so--called amplitude equation is derived 
and it is shown that the bifurcation to microtubule oscillations
is supercritical.

\end{abstract}
 %
% ****************************************************************

%   PACS

% ****************************************************************

%

\begin{multicols}{2}

\narrowtext
%
% ****************************************************************
%   Table of contents (remove this before sending!)
% ****************************************************************
%
%\tableofcontents
%
% ****************************************************************
%   Chapters 
% ****************************************************************
%

\section{Introduction}
\label{secintro}
Microtubules  are  cylindric filaments that are
used in  cells for many  different purposes, being vitally
involved in cell motility and division, in organelle
transport, and in cell morphogenesis and organization \cite{Alberts:94}. 
The precise ways in which microtubules achieve their amazing variety of 
cellular functions is not fully understood yet. 
Microtubules in cells are generally dynamic, they assemble,  disassemble
or rearrange on a time scale of minutes.
GTP (guanosine triphosphate) hydrolysis is apparently the 
driving force of microtubule physiology.

The rich non--equilibrium dynamics of microtubules,
 including the nucleation and 
polymerization kinetics etc.  \cite{Mitchison:97.1,Job:01.1}
 is attracting considerable 
attention, both experimentally and theoretically
\cite{Mitchison:84.1,Pirolet:87.1,Hill:87.1,Mandelkow:88.1,Mandelkow:88.2,Melki:88.1,Mandelkow:89.1,Job:89.1,Mandelkow:92.1,Hill:87.2,Leibler:93.1,Mandelkow:94.1,Houchman:96.1,Flyvbjerg:97.1,Sept:99.2}.
Two phenomena in this area, the dynamical instability of microtubules
 \cite{Mitchison:84.1} and 
the oscillatory polymerization 
\cite{Pirolet:87.1,Hill:87.1,Mandelkow:88.1,Mandelkow:88.2,Melki:88.1,Mandelkow:89.1,Job:89.1,Mandelkow:92.1}
challenge theoretical modeling already for a while
\cite{Hill:87.2,Leibler:93.1,Mandelkow:94.1,Houchman:96.1,Flyvbjerg:97.1,Sept:99.2}.

Oscillations during  microtubule
polymerization have been  observed  either when   GTP is 
 regenerated enzymatically from endogenous GDP (guanosine diphosphate)
\cite{Pirolet:87.1,Mandelkow:88.1,Melki:88.1,Job:89.1} or when 
some amount of GTP  is provided at the beginning 
 or during an experiment.  
In the latter case oscillations occur only as a transient, because
GTP is either consumed or some reactions steps may be
inhibited due to the   accumulation of GDP \cite{Mandelkow:88.2}.
 If both possibilities are combined, the length
of   a  transient regime
depends  on the
initial concentrations of GTP and GDP  and on  
 the  capacity   to regenerate   GTP. 
Present models for microtubule polymerization 
focus mainly on a description of 
  transiently occurring  oscillations and 
the solutions of  the respective  models 
are mostly numerical \cite{Hill:87.1,Mandelkow:94.1,Flyvbjerg:97.1}.

In recent in vitro experiments,  however,  the  capacity to regenerate 
 GTP  has been 
enhanced and extended   up to several hours  \cite{Tabony:92.1}.
Compared to a typical  oscillation period during  microtubule polymerization, which
is of the  order of  a minute, the reaction conditions in these  experiments 
are almost quasi--stationary 
over  a long range of time.  Therefore we 
focus on modeling microtubule polymerization 
for    time--independent  regeneration conditions. As starting point 
we take
 common reduced  models, where several
elementary processes of the real biochemical 
reaction are described by a  few
effective reaction steps as explained in Sec.~\ref{smodel}, 
cf. Refs.~\cite{Hill:87.1,Mandelkow:88.1,Mandelkow:94.1,Flyvbjerg:97.1}.
Reductions of complex chemical reaction schemes are quite common and 
a famous  example is the   so--called oregonator \cite{Scott:94} that  is  
a reduced model for the legendary Belousov--Zhabotinsky (BZ) reaction
\cite{Tyson:76}. 
However, since  microtubules are long filaments, there are
essential  differences between  the
polymerization of microtubules filaments 
 and common  chemical reactions.
For instance microtubules may undergo an orientational ordering transition 
beyond a critical filament density 
\cite{Hitt:90.1}, 
a phenomenon,  which doesn't    occur  in common  chemical reactions.
Accordingly, the length distribution of microtubules
is explicitly taken into account for all variants of 
models investigated in this work.
Such models are the basis of future work on  
  interesting 
pattern--formation  phenomena related to the interplay between  
orientational  ordering of the filaments 
 and the kinetics involved in the filament 
growth  \cite{Ziebert:02.120}.

In addition,   
we focus on model variants  that 
include the possibility of an 
 oscillatory microtubule polymerization and that allow
analytical approaches. However, the reaction 
steps, such as nucleation, growth and decay of microtubules or the rate limiting factors 
of oligomer decay or tubulin regeneration, which  have been identified to be crucial
for oscillations 
 \cite{Pirolet:87.1,Hill:87.1,Mandelkow:88.1,Mandelkow:88.2,Job:89.1},
 are taken into account. Moreover  we address  
 the question whether  microtubule oscillations occur transiently or in a persistent manner 
 beyond a   Hopf bifurcation.  Whether 
such a  Hopf bifurcation takes  place super- 
or subcritically is  investigated
in terms of the  
so--called amplitude expansion.

It is not a major goal of this work to achieve 
quantitative agreement between the results 
obtained with  phenomenological models 
and  experimental measurements. However,
since the present understanding of the mechanism leading to
oscillatory microtubule polymerization is incomplete,  
reduced models 
may be  an appropriate tool for working out   
typical trends that may be  testable in experiments.
For comparison it is very
helpful 
 that for reduced models trends  may  
be worked out  analytically and  may
 be  presented by 
simple formulae. A  number of spatiotemporal phenomena 
involving microtubule polymerization 
  call for a better understanding too
\cite{Mandelkow:89.1,Tabony:90.1,Leibler:97.1}, but 
also in this case simple and effective  models are indispensable in order 
to keep the modeling tractable \cite{Hammele:02.120}.

At the  transition to  oscillatory
polymerization  the stationary state
becomes sensitive against small  perturbations, which   grow or decay  exponentially,
  $\propto e^{\sigma t}$. Here    the exponential factor  $\sigma =  \sigma_r \pm i \omega_c $ 
is the sum of the so--called growth rate $\sigma_r$ and
the oscillation frequency $\omega_c \not = 0$.
Below the bifurcation point
the growth rate 
$\sigma_r <0$ is negative and the perturbations
are damped.  Beyond the bifurcation point 
  $\sigma_r$ is positive and the stationary polymerization state
is  unstable against  oscillatory perturbations. Hence 
the  Hopf bifurcation to oscillatory polymerization   takes place 
when  the real part  $\sigma_r$ 
of both  roots passes zero. The investigation of the  
polymerization dynamics beyond the Hopf bifurcation  
requires in most cases   a numerical analysis of the basic reaction equations. 
However, close to  threshold  $\sigma_r$ is small 
and 
the oscillation  of the polymerization, described by the
real part of 
  $e^{i \omega_c t}$, is 
much faster than the  temporal evolution 
of the complex valued amplitude $A(t)$  of the
oscillations.
Therefore the  oscillation may  be written
as a product of both factors, i.e.  
$\propto A(t) e^{i \omega_c t}$, and there is 
a very general approach, the so--called amplitude expansion,
for separating the dynamics at these
two disparate time scales  \cite{CrossHo,Strogatz:94}.
The  amplitude equation describing  the evolution of the amplitude $A(t)$
is obtained by  a perturbation expansion of the 
reaction equations with respect to the slowly varying amplitude $A(t)$
and it is of the form
\begin{equation}
\label{ampliin}
\tau_0 \partial_t A = \varepsilon ( 1 + i a) A  - g(1+ic) \mid A\mid^2 A\,.
\end{equation}
The control parameter $\varepsilon$  
measures  the relative distance from the
bifurcation point and $\tau_0$ is the relaxation time defined  
by $\tau_0 = \varepsilon / \sigma_r$,
 that  depends on the system.
If the coefficient $g$ of the nonlinear term is  positive, the
bifurcation to the  oscillatory state is supercritical and if 
it is negative, the bifurcation is  subcritical. 
The imaginary parts of the prefactors 
 describe the linear and the nonlinear frequency dispersion. 
Especially about the extension  Eq.~(\ref{ampliin}) including  
spatial degrees of freedom, there exists 
  a rich   literature as  summarized e.g. in a
recent review   \cite{Aranson:02.1}.
Here in this work we calculate  the coefficients of the  universal equation
(\ref{ampliin}) 
for microtubule polymerization 
and we discuss their variation in terms of the reaction rates.

In the following  section \ref{smodel} we describe the main steps of the reaction cycle for
microtubule polymerization and the respective
equations for two models are presented. 
The  time--independent solutions for the stationary 
polymerization  are
given for both models  analytically in 
section \ref{sstasol}. Those become unstable against
oscillatory perturbations
in the  range of  high  of  tubulin--dimer density.
The respective linear stability analysis and the derivation of the 
oscillation  threshold   are given 
 in section  \ref{sthreshosz}, including their 
dependence on the reaction parameters. 
Readers who are mainly interested in numerical results  
about the oscillation threshold 
may  proceed directly to section
\ref{resthreshI}.
 The partial 
differential equations for  growing and shrinking  microtubules are
of first order in the length microtubules  and  first order in time. 
Their  straight forward discretization 
and numerical solution  has to be considered with care, therefore 
 a stable numerical
scheme, that becomes exact close to 
the threshold of the Hopf bifurcation, is described 
in section \ref{snumeric}. The derivation of
the universal equation (\ref{ampliin}) is outlined  in  Sec.~\ref{sampli}
whereas the technical details are given in appendix~\ref{appA}.
With a summary and an outlook about 
modeling microtubule polymerization we conclude this work in  Sec.~\ref{summary}.

%%%%%%%%%%%%%%%%%%%%%%%%%%%%%%%%%%%
\section{Models for microtubule polymerization}            %
\label{smodel}
%%%%%%%%%%%%%%%%%%%%%%%%%%%%%%%%%%%

Microtubule assembly  and disassembly proceeds in several steps
 \cite{Alberts:94,Mitchison:97.1,Job:01.1,Pirolet:87.1,Mandelkow:88.1,Mandelkow:88.2}.
Aggregation of {\it guanosine triphosphate} (GTP) liganded tubulin dimers,
the so--called tubulin--t, to microtubules
is started by heating up tubulin solutions 
to a temperature of about $30-37^o$C in the presence of GTP.
Then microtubules  spontaneously nucleate and  
polymerize to long rigid polymers made
up of $\alpha-\beta$ tubulin dimers.
An increasing number of 
long microtubules in a solvent  causes  an increasing turbidity and the  
amount of polymerized tubulin--t may be monitored
by measuring  this turbidity \cite{Hill:87.1} 
or  by X-ray scattering \cite{Mandelkow:88.2}. 
The nucleation of microtubules is a rather  complex 
process and it is still  a matter of debate whether 
the nucleation rate depends in experiments only   on the
initial concentration of tubulin-t, $c_t$,  
 or during the polymerization on the 
temporally varying  $c_t$   \cite{Job:01.1,Job:00.1}. 
But once microtubules are formed, 
they  grow  and  the
available tubulin--t dimers will be used up. 
The growth velocity of microtubules, $v_g$,
is rather sensitive to temperature variations  
 but it is rather  independent on $c_t$ \cite{Job:00.1,Libchaber:94.1}.
Growing microtubules may change their
state to rapidly depolymerizing ones
by  the so--called {\it catastrophe} rate $f_{cat}$.
 In previous works for the catastrophe rate 
mostly  an exponential dependence   on the tubulin--t concentration
was assumed, i.e. 
   $f_{cat}\sim \exp(-c_t/c_f)$ with some constant $c_f$
\cite{Hill:87.1,Mandelkow:94.1}.
Once  microtubules have changed from growth to shrinking, they shrink rather quickly with 
a large velocity $v_s \gg v_g$.

During the depolymerization of 
microtubules they are fragmented into oligomers or
directly into 
 {\it guanosine diphosphate} (GDP) liganded tubulin dimers,
the so--called tubulin-d dimers. 
The oligomers themselves  are believed to fragment further 
into tubulin--d dimers and the decay rate depends on the 
free GTP and GDP. Oligomers are stabilized by GDP and
destabilized by  GTP \cite{Mandelkow:88.2,Job:89.1}.
If an  excess of GTP is available, then 
tubulin--d in solution
will exchange its unit of GDP for GTP and  each tubulin--t dimer 
resulting from such an  exchange step  is identical to the initial
tubulin--t dimer. Such a regeneration step completes the whole 
microtubule polymerization  cycle.
If a continuous source of GTP is provided, for instance by a regeneration process, 
 this cycling  may be continued over a long time \cite{Tabony:92.1}.
The variation  of the reaction rates of the 
polymerization cycle  with  the concentration  $c_t$
may depend on the specific experiment.

% ******************************************************************************
%   Figure 1
% ******************************************************************************
%
\begin{figure}[htb]
\epsfxsize 8.8cm \ifpreprintsty \epsfxsize 15.0cm \fi
\noindent
\epsfbox{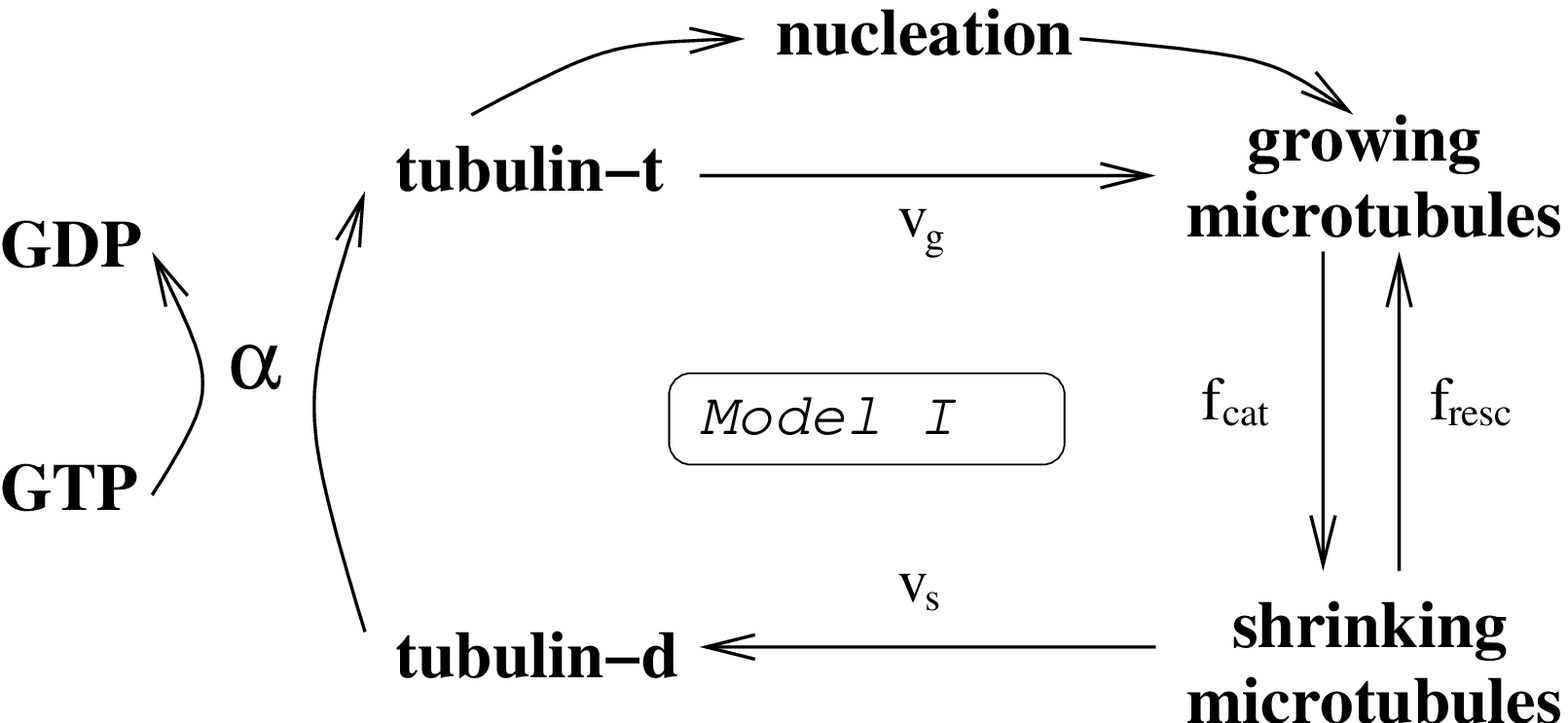}
\vspace{1.0cm}
\\
\noindent
\vspace{0.8cm}
\epsfxsize 8.8cm \ifpreprintsty \epsfxsize 15.0cm \fi
\epsfbox{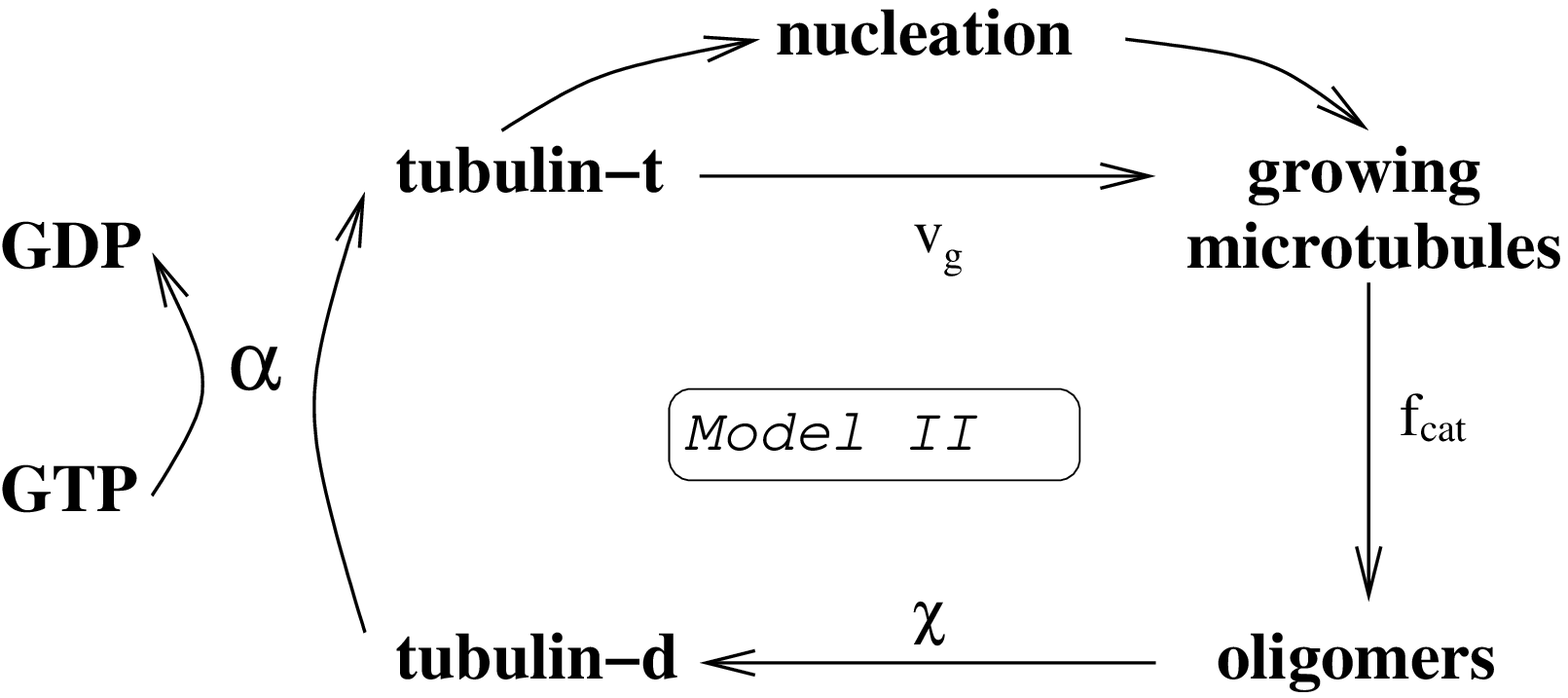}
\caption [Figure 0a] 
  {Two models for the cycle of microtubule polymerization. 
Model I (upper cycle): Tubulin--t dimers may spontaneously form
nuclei of microtubule  that  grow further by incorporating 
 tubulin--t dimers. A growing microtubule may also 
change its state to a quickly  depolymerizing one by the so--called
{\it catastrophe} rate $f_{cat}$, but it may also change back to the 
polymerizing state by a so--called and  rather small {\it rescue} rate $f_{resc}$.
Tubulin--d dimers are released during this  microtubule depolymerization
 and the  whole cycle becomes closed 
by regenerating them by a rate $\alpha$ back to tubulin--t dimers.
Model II (lower cycle): Here the intermediate step of shrinking microtubules is replaced by
oligomers e.g. microtubule break off 
with a rate  $f_{cat}$ directly into oligomers and the oligomers themselves  may
break off with the rate  $\chi$ into tubulin--d dimers. 
The rest of the cycle is identical with the  upper cycle.
}
\label{figm1}
\end{figure}
%*****************************************************************************************

There are rather detailed models available 
to describe this reaction cycle of microtubule polymerization, see e.g. \cite{Mandelkow:94.1}.  
As a simplification of this complex  biochemical reaction  we  only take into account 
as rate limiting factors 
one of the  two  intermediate steps of the polymerization cycle, either the dynamics of  
shrinking microtubules (upper cycle in Fig.~\ref{figm1}) 
or the decay dynamics of oligomers   (lower cycle in Fig.~\ref{figm1}). 
Without both rate rate limiting factors there are no microtubule oscillations,
but one is sufficient for oscillations. The two simplified  reaction schemes, as 
sketched in Fig.~\ref{figm1},  are analyzed in detail in this work.

%------------------------------------------------
\subsection{Dynamics of growing microtubules}
\label{dynmicro}
%------------------------------------------------

During  microtubule polymerization 
there are  many growing filaments in a unit volume and their 
length distribution may be 
described by a length and time--dependent function 
$p_g(l,t)$ whose detailed form  varies 
with the  experimental conditions. A
simple model for the  dynamics of distribution of growing 
 microtubules
$p_g(l,t)$ is described by  the following first 
order differential equation 
 \cite{Hill:87.2,Leibler:93.1}
\begin{eqnarray}
\label{pgl1}
\partial_{t}p_{g} &=& - f_{cat} \, p_{g}  
- v_{g} \frac{\partial p_{g}}{\partial l}\,. 
\end{eqnarray}
 $f_{cat}$ describes either 
 the transition  from the  growing to the shrinking state of   
microtubules  (model I) 
 or the decay of  
growing microtubules
into oligomers (model II). $v_g$ is the growth velocity of the microtubules. 

%----------------------------------------------------
\subsubsection{Growth velocity and catastrophe rate}
%----------------------------------------------------

In recent experiments  with  high 
tubulin--t concentration the  
growth velocity $v_g$  was rather independent of $c_t$    \cite{Job:00.1}. 
Since we are mainly interested in the oscillatory
behavior of microtubule polymerization, that occurs 
at high   $c_t$--concentrations, we assume 
a constant $v_g$ in this work. 
In most of the present  models   a $c_t$--dependent 
catastrophe rate $f_{cat}$ 
is  crucial for oscillatory polymerization of  
microtubules.   Rather common is 
an exponential $c_t$--dependence
\cite{Mandelkow:94.1}  

\begin{eqnarray}
\label{rateexp}
 f_{cat}(c_t) &=& f  \,\,e^{- c_t/c_f} \,, 
\end{eqnarray}
with the amplitude $f$ and the decay constant $c_f$.  
However, also  a  linear $c_t$--dependence 
\begin{eqnarray}
\label{ratelin}
 f_{cat}(c_t) &=& \bar f \,( c_u - c_t) \,, 
\end{eqnarray}
with an appropriate constant $c_u > c_t$ 
leads to an oscillatory microtubule polymerization as we show
in Sec.~\ref{osthreshI}. 
A hyperbolic $c_t$--dependence  of $f_{cat}$, as discussed in  
Ref.~\cite{Houchman:96.1}, also supports oscillating
polymerization.  

%--------------------------------------------------
\subsubsection{Nucleation and boundary conditions}
%--------------------------------------------------

The nucleation process of microtubules is rather complex and it 
has been  investigated  in more detail in 
Refs. \cite{Mandelkow:92.1,Libchaber:94.1,Flyvbjerg:96.1}, recently.  
The nucleation rate $\nu$  depends on the initial concentration $c_0$ of 
tubulin dimers,   but  it is  rather independent 
of  the temporal variation of $c_t$,
as observed in recent  experiments
 \cite{Job:00.1,Job:01.1}.
Accordingly, for a given initial 
concentration $c_0$  we assume 
a constant  nucleation rate $\nu$. The nucleation rate 
$\nu$ itself  defines a
boundary condition for the length distribution
of growing microtubules 
$p_g(l,t)$ at $l=0$,
\begin{equation}
\label{boundpg}
p_g (l=0,t) = {\nu \over  v_g} \,.
\end{equation}
%

%--------------------------------------------------------------------
\subsection{Model I includes the dynamics of shrinking microtubules} 
%--------------------------------------------------------------------

In model I we take into account 
as  an intermediate step between 
 growing microtubules and tubulin--d dimers the dynamics  of 
 shrinking microtubules, $p_s(l,t)$. Here 
the catastrophe rate  $f_{cat}$  describes the transition 
of microtubules from the 
growing to the shrinking state.
 The depolymerization speed $v_s$ of shrinking microtubules, $p_s(l,t)$,
is mostly  much larger than the growth velocity $v_g$. 
Having microtubules in two different states, one may also 
expect  a transition from the shrinking back to
the growing state, as described by a rate $f_{resc}$. 
So one has two coupled equations for the
growing and shrinking microtubules \cite{Hill:87.2,Leibler:93.1}
\begin{mathletters}
\label{pgl}
\begin{eqnarray}
\label{pgl1I}
\partial_{t}p_{g} &=& - f_{cat} \, p_{g} + f_{resc} \, p_{s} 
- v_{g}\partial_{l}p_{g} 
%+ D(l) \Delta p_{g} 
\, , \\
\label{pgl2I}
\partial_{t}p_{s} &=&  f_{cat}\, p_{g} - f_{resc}\,  p_{s} 
+ v_{s}\partial_{l}p_{s}  
%+  D(l) \Delta  p_{s}  
\,.
\end{eqnarray}
\end{mathletters}
The rescue rate $f_{resc}$,  however,  
is usually very small in experiments and therefore  
it is neglected in this work. 
The boundary condition for shrinking 
microtubules is 
\begin{equation}
\label{boundps}
p_s (l \to \infty,t) = 0 \,,
\end{equation}
because the transition from  growing to shrinking 
microtubules is  the only
source for the shrinking ones and $p_g( l \to \infty,t)$ 
vanishes for large values of $l$.

The temporal evolution of the concentration of the 
tubulin--t dimers $c_t$ and tubulin--d dimers $c_d$
is  described by two  equations as follows 
\begin{mathletters}
\label{ctcd}
\begin{eqnarray}
\label{ctcd1}
\partial_{t}c_{t} &=& -\gamma v_{g}\int_{0}^{\infty}dl \, p_{g}(l,t) + \alpha c_{d} 
%+ D_c \Delta  c_{t}
\,,
\\
\label{ctcd2}
\partial_{t}c_{d} &=& \gamma v_{s}\int_{0}^{\infty} dl \, p_{s}(l,t) - \alpha c_{d} 
%+ D_c \Delta  c_{d}
\, .
\end{eqnarray}
\end{mathletters}
The first term in Eq.~(\ref{ctcd1})
describes the consumption of tubulin--t during the
polymerization (growth) of microtubules and $\gamma$ is a length
factor describing the number of tubulin dimers that are incorporated in 
a unit length of microtubules.
 $c_t$ is regenerated  from 
 $c_d$  by exchange the unit  GDP for GTP and  this regeneration process, 
described by the rate 
$\alpha$, occurs in Eq.~(\ref{ctcd1}) as a source and in 
Eq.~(\ref{ctcd2}) as a sink.
Tubulin--d dimers are released during the
depolymerization  of microtubules $p_s(l,t)$ and this source   is described 
by the integral in Eq.~(\ref{ctcd2}).

Tubulin dimers may be a constituent of  growing or shrinking 
microtubules or they  carry GTP or GDP as single dimers,
but altogether they are conserved as 
 expressed by the condition 
\begin{equation}
\label{erhalt}
  c_{t} + c_{d} + \gamma L = c_{0}\, .
\end{equation}
Here $c_0$ describes the  overall  concentration of tubulin dimers and
$L(t)$ is the integrated   length of all microtubules per unit volume
\begin{equation}
\label{conserv}
  L(t) = \int_{0}^{\infty} dl \,\,   l \Big(p_{g}(l,t) + p_{s}(l,t)\Big )  \,.
\end{equation}
The tubulin--d concentration  $c_d$ may be 
eliminated from  Eq.~(\ref{ctcd1}) by using the
conservation condition (\ref{erhalt}).
On the other hand Eq.~(\ref{erhalt}) in combination with 
Eqs.~(\ref{pgl}) and  Eq.~(\ref{ctcd1})  yield   an 
equation that is identical with  Eq.~(\ref{ctcd2}).
Hence Eqs.~(\ref{pgl1I}) and (\ref{pgl2I}) together with 
\begin{eqnarray}
\label{ceqsh}
\partial_{t}{c}_{t} &=& - \gamma \int_{0}^{\infty} dl \, \Big( v_{g} p_{g} 
+ \alpha  l( p_{g}+  p_{s}) \Big)  + \alpha (c_{0} - c_{t})\, ,
\end{eqnarray}
describe the polymerization dynamics  of microtubules
for model I,  whereby a 
constant growth and shrinking velocity
is assumed in this work.

%%%%%%%%%%%%%%%%%%%%%%%%%%%%%%%%%%%%%%
\subsubsection{Rescaling of model I}
%%%%%%%%%%%%%%%%%%%%%%%%%%%%%%%%%%%%%%
After rescaling time $t$ and length $l$, i.e.
\begin{eqnarray}
t' = \alpha t\,, \qquad  l' =  \frac{ \alpha }{v_g}l,
\end{eqnarray}
it is easy to see that model I  may be characterized
by a  set of dimensionless parameters 
\begin{eqnarray}
 \quad \gamma\frac{ v_g}{\alpha}, \quad 
\frac{ v_s}{v_g}\,,\quad \frac{\nu}{\alpha}\,,\quad \frac{c_0}{c_f}\,,\quad
\frac{ f_{resc}}{\alpha}\,.
\end{eqnarray}
Some of these dimensionless quantities 
 may be further combined 
to other dimensionless parameters 
as for instance in the  threshold condition
given in  
Sec.~\ref{osthreshI}.

%
%
%----------------------------------------------
\subsubsection{Reduced model} 
%-----------------------------------------------
\label{modred}
Since the  depolymerization  velocity $v_s$  is much larger 
than the growth  velocity  $v_g$ one may
also consider  the limit   $v_s\gg v_g$. In this case  
the  shrinking microtubules decompose nearly 
instantaneously into tubulin--d dimers 
and  growing microtubules decay effectively, due to the short 
life time of the shrinking microtubules,  
into tubulin--d dimers. In order to describe this
direct decay the source term 
in Eq.~(\ref{ctcd2}), $ \gamma v_{s}\int_{0}^{\infty}dl\, p_{s}(l,t)$,
must be replaced by  $ \gamma f_{cat} \int_{0}^{\infty} dl\,l \,p_{g}(l,t)$.
Eliminating again the density  $c_d$ one ends up with a reduced  model 
for only two densities:
\begin{mathletters}
\label{redmod}
\begin{eqnarray}
\label{pglsh}
\partial_{t}p_{g} &=& - f_{cat} \, p_{g} 
- v_{g}\partial_{l} p_{g}   \,,\\   
\label{ceqshr}
\partial_{t}{c}_{t} &=& - \gamma \int_{0}^{\infty} dl \, \Big( v_{g}  
+ \alpha  l  \Big)p_{g}   + \alpha(c_{0} - c_{t})\,.
\end{eqnarray}
\end{mathletters}
This simplified model reproduces essential aspects  of stationary 
 polymerization of microtubules  as described in Sec.~\ref{sstasol}. 

%-----------------------------------------
\subsection{Model II includes the dynamics of oligomers}
%-----------------------------------------
Oligomers occur  as an intermediate product during the
decay of microtubules and they are made of 
 several tubulin dimers. This intermediate product is 
ignored   in model I. Here  in model II, after the so--called
catastrophe, we ignore       the
dynamics of shrinking microtubules as an intermediate step
and instead we take into account 
the (decay) dynamics of oligomers. 
Therefore, the  catastrophe rate 
$f_{cat}$ in Eq.~(\ref{pgl1}) 
describes for  model II a
direct transition of  growing microtubules
into oligomers. Furthermore, it is  assumed that 
oligomers  decay 
with  the rate $\chi$ into tubulin--d dimers.
The  concentration of oligomers is denoted by 
$c_{oli}$ and  its dynamics as well as that of $c_d$ are described by the two equations
\begin{mathletters}
\begin{eqnarray}
\label{olico}
\partial_t c_{oli} &=& \eta f_{cat} \int_0^{\infty} dl \, l\, p_g(l,t) 
                             - \chi c_{oli} \, ,\\
\label{olicd}
\partial_t c_d &=& \chi\lambda \, c_{oli} - \alpha c_d \, .
\end{eqnarray}
\end{mathletters}
$\eta$ is a measure for the number of oligomers 
per unit length of the microtubules and 
$\lambda$ is a measure for the number
of tubulin dimers per  oligomer.
Oligomers decaying  with the rate $\chi$ build  a source term 
in the equation for tubulin--d dimers in Eq.~(\ref{olicd}).

The conservation law for the  concentration of
all tubulin dimers  takes the form
\begin{eqnarray}
\label{olierh}
c_t + c_d + \lambda c_{oli} + \eta \lambda  \int_{0}^{\infty} 
\, dl \, \,l \, p_g(l,t) = c_0 \, .
\end{eqnarray}
The equation for the  growing microtubules is the same as
for model I, cf.  Eq.~(\ref{pgl1}), but in the equation  for  $c_t$, cf.  
Eq.~(\ref{ctcd1}),  one has to replace the 
length factor $\gamma$ by the product $\eta \lambda$. 
Eliminating $c_{oli}$ model II is described by 
Eq.~(\ref{pgl1}) and Eq.~(\ref{ctcd1}) together with
the following  dynamical equation for $c_d$ 
\begin{eqnarray}
\label{cderh}
\partial_t c_d &=& \chi \left( c_0 - c_t - c_d 
                       -  \eta \lambda \int_0^{\infty} dl \, \,l\, p_g \right) 
                       - \alpha c_d  \, .
\end{eqnarray}
As  boundary condition for the growing microtubules
we again  use 
Eq.~(\ref{boundpg}) with  a constant nucleation rate $\nu$.
For model II we only consider the  catastrophe rate
given  in Eq.~(\ref{rateexp}). This again
guarantees a nonlinear feedback of the dynamics of the tubulin--t dimers to
the dynamics of the growing microtubules.

%------------------------------------
\subsubsection{Reduced model}
%------------------------------------

Similar as for model I also model II becomes 
 in the limit $\chi \to \infty$  
identical with  the model  described by 
Eqs.~(\ref{redmod}). If we assume 
a very fast  dissociation of oligomers into tubulin--d dimers,
  $\chi \gg 1$, we can neglect
the intermediate state $c_{oli}$.
In this case the source term in equation (\ref{olicd}),
$\chi \lambda \,c_{oli}$, can be  replaced by the source in equation 
(\ref{olico}), cf.
$\eta \lambda f_{cat} \int_0^{\infty}\,dl \, l \, p_g $,
which describes the direct decay of growing microtubules
into tubulin--d dimers.
After replacing $c_d$ and  setting $\gamma=\eta \lambda$ in 
Eq.~(\ref{ctcd1}), 
we again obtain
with the help
of the conservation law (\ref{olierh}) 
the simple reduced model as described by the equations 
(\ref{redmod}). 

%%%%%%%%%%%%%%%%%%%%%%%%%%%%%%%%%%%
\section{Stationary solutions}     %
\label{sstasol}
%%%%%%%%%%%%%%%%%%%%%%%%%%%%%%%%%%%%

A  polymerization cycle with a stationary 
length distribution of
microtubules and time--independent dimer concentrations $c_{t}$, $c_{d}$  
or oligomer concentration $c_{oli}$
are one type of  the solutions of the
model equations described in the previous section
\ref{smodel}. For this stationary state 
the various polymerization steps,
such as nucleation, assembly and disassembly
of microtubules as well as the regeneration of tubulin-d dimers
are in a balanced state. Under certain conditions a 
stationary polymerization 
is observed in 
experiments 
 \cite{Libchaber:94.1}. However,  it
may become unstable against
oscillatory perturbations  if the initial
tubulin dimer concentration  
$c_0$ is large enough, as 
shown  in the following section
\ref{sthreshosz}.

%-----------------------------------
\subsection{Model I }
%-----------------------------------

Eqs.~(\ref{pgl}) are
 first order  linear differential equations  
with respect to the length $l$ and in the stationary case these equations 
 have  exponentially decaying  solutions, which take in the limit of
a vanishing rescue rate the form 
\begin{eqnarray}
\label{pgsstat}
  p_{g,s}^{(0)} (l) &=& \frac{\nu}{v_{g,s}} \exp{\left(-\frac{f_{cat}^{(0)}}{v_{g}}l \right)} \,.
\end{eqnarray}
The catastrophe rate $f_{{\it cat}}^{(0)}$ may be given either
by Eq.~(\ref{rateexp}) or Eq.~(\ref{ratelin})  but
in both cases the stationary
 tubulin--t concentration, denoted by  
 $c_t^{(0)} $, is determined 
self--consistently as described below.
The stationary solutions  
$p_{g,s}^{(0)}$  allow an  analytical calculation
of  the integrals in Eq.~(\ref{ceqsh}) and  a nonlinear
equation in $c_t^{(0)}$  follows,
\begin{eqnarray}
\label{cgleich0}
c_{0} - c_t^{(0)}  &=& { \nu \gamma v_g \over f_{cat}^{(0)} }  
 \left( \frac{1}{\alpha}   +
{1 \over f_{cat}^{(0)} } ( 1 + \beta )\right) \,\, .
\end{eqnarray}
From  this equation 
the stationary tubulin concentration   $c_t^{(0)}$ 
can be determined as a function
of the overall concentration of tubulin dimers $c_0$ and 
as function of the other  parameters. In 
Eq.~(\ref{cgleich0})
 the abbreviation for  the velocity ratio
\begin{equation}
  \label{eq:beta}
  \beta = \frac{v_g  }{v_s }
\end{equation}
has been introduced and  the respective  length distributions 
$p_{g}^{(0)}$  and $p_{s}^{(0)}$ 
follow for a given value of $c_t^{(0)}$
via Eqs.~(\ref{pgsstat}).  
The stationary value $\ct{0}$
for the reduced model, described by  Eqs.~(\ref{redmod}),  follows
from  Eq.~(\ref{cgleich0}) in the limit  $ \beta \to 0$.

% ******************************************************************************
%   Figure 2
% ******************************************************************************
% 
\begin{figure}[htp] 
\vspace{-0.4cm}
\epsfxsize 8.7cm \ifpreprintsty \epsfxsize 15.0cm \fi
\epsfbox {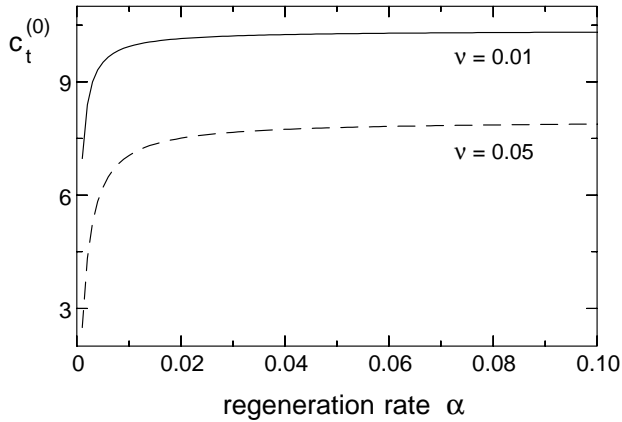}
\caption [Figure 0a] 
  {The
 tubulin--t concentration $c_t^{(0)}$ 
for the stationary polymerization state of model I 
is shown as a function of the regeneration rate
$\alpha$ and  for two different values of the nucleation
rate $\nu$. The velocity ratio between  the growing and shrinking 
microtubules is $\beta = v_g/v_s = 0.1$
and the  rest of  parameters are
$c_0=120, v_g=0.1, c_f=3, f=0.1, \gamma=1$.
}
\label{fig0a}
\end{figure}
%********************************************************************************

In the range of $\alpha$ much larger than the catastrophe rate $\fcat$
the stationary tubulin--t 
concentration $c_t^{(0)}$ becomes independent of it,
because  all tubulin--d dimers, that are 
released during
the depolymerization, are immediately regenerated to tubulin--t dimers.
Both  a large nucleation rate $\nu$ and a large 
growth velocity $v_g$ lead to a high consumption of tubulin--t 
and therefore to a lower
stationary concentration  $c_t^{(0)}$. This tendency is 
illustrated by the difference between the two 
curves in Figure \ref{fig0a}.
On the other hand, large values of the amplitude
of the respective  catastrophe rate, either  
$f$ or $\bar f$,  act against long 
microtubules which consist  of many  tubulin dimers   and therefore enhance 
the densities  $c_t^{(0)}$ and  $c_d^{(0)}$.
The length distribution of the growing respective shrinking
microtubules is determined by  $v_g/\fcat$, which also depends via the
catastrophe rate on the concentration  $c_t^{(0)}$.

These  tendencies becomes even more obvious  
 if the linear dependence of the 
catastrophe rate in   Eq.~(\ref{ratelin}) is chosen
for the special case  $c_u=c_0$ and 
Eq. (\ref{cgleich0}) is expanded in the limit of small 
and large values of $\alpha$.
In  both  cases we obtain the simple 
formulas 
\begin{mathletters}
\begin{eqnarray}
c_t^{(0)} &=& c_0 - \left( \frac{\nu \gamma v_g\, (1+\beta)}{\bar f}\right)^{1/3} \quad
( \alpha \gg \bar f) \,,\\
c_t^{(0)} &=& c_0 - \left( \frac{\nu \gamma v_g\, }{\alpha \bar f}\right)^{1/2} 
\hspace{16mm} (\alpha \ll \bar f)\, ,
\end{eqnarray}
\end{mathletters}
which reflect the described tendencies.

%
%Using instead of a concentration dependent
%catastrophe rate 
%a constant one, $f_{cat}=const.$  and 
%a tubulin dependent polymerization velocity, cf.
%$v_g = \tilde v_g c_t$, leads also to a coupling between
%the    density $c_t^{(0)}$
%and therefore an equilibrium distribution of microtubule $p_{g,s}^{(0)}$. 
%
%

%-----------------------------------
\subsection{Model II}
%-----------------------------------

Stationary solutions for model II can be calculated
in a similar manner as discussed in the previous section for model I.  The length distribution
of  growing microtubules 
is again given by Eq.~(\ref{pgsstat}) and
the integral in Eq.~(\ref{cderh}) can  be calculated
analytically.   $c_d$ may be  eliminated from  Eq.~(\ref{cderh})
by using Eq.~(\ref{ctcd1}) with $\dot c_t=0$ and 
by setting $\gamma=\eta \lambda$.
Then
the  nonlinear equation for tubulin--t dimers $\ct{0}$ takes the form 
\begin{eqnarray}
\label{statoli}
c_0 - \ct{0} = \frac{\nu \eta \lambda v_g}{\fcat} \left( \frac{1}{\alpha} + \frac{1}{\chi} 
                        + \frac{1}{\fcat} \right ) \,.
\end{eqnarray}
Eq.~(\ref{statoli}) is invariant under permutation 
$\alpha \leftrightarrow \chi$.  
If  $\alpha$ and $\chi$ become much larger than the
catastrophe rate, the stationary concentration $\ct{0}$
becomes  rather independent of both. 
In the limits $v_s \to \infty$ in Eq.~(\ref{cgleich0})
and $\chi \to \infty$ in Eq.~(\ref{statoli})
we again obtain the  concentration $\ct{0}$ for the reduced model.
No stationary  solution is possible in the limit
$\chi \to 0$ because in this limit all tubulin--d
dimers are stored in oligomers and the polymerization cycle 
becomes interrupted.

%%%%%%%%%%%%%%%%%%%%%%%%%%%%%%%%%%%%
\section{Threshold for oscillatory polymerization}      %
%%%%%%%%%%%%%%%%%%%%%%%%%%%%%%%%%%%%
\label{sthreshosz}

Stationary  microtubule polymerization
 becomes unstable against  oscillating 
modes in the range of  high tubulin dimer concentrations 
$c_0$ and   the parameter range where  this happens  
is calculated  by a linear stability analysis. 
Starting from the model equations given in  Sec.~\ref{smodel} 
we derive linear equations for
 small perturbations with respect to the   stationary
state and such perturbations  exhibit an exponential time dependence,
  $ e^{\sigma t}$. For the exponential factor 
  $\sigma$ we derive a   nonlinear equation 
from which both   
the critical dimer concentration $c_{0c}$ and the critical
frequency $\omega_c$  for the Hopf bifurcation
is  calculated numerically for various parameter combinations 
and  in limiting cases also analytically.

%---------------------------------------------
\subsection{Model I}
\label{osthreshI}
%---------------------------------------------

We introduce  small perturbations  $\p{g,s}{1}$ and $\ct{1}$ with
respect to the stationary solutions 
 $\p{g,s}{0}$ and $\ct{0}$ as determined in the last section. 
With the ansatz 
\begin{mathletters}
\begin{eqnarray}
p_{g,s} &=& p_{g,s}^{(0)} +p_{g,s}^{(1)} \, , \\
c_t &=& c_t^{(0)} +c_t^{(1)}\, ,
\end{eqnarray}
\end{mathletters}
one obtains, after
linearization of  Eqs.~(\ref{pgl}) and Eq.~(\ref{ceqsh}), 
the following set of linear equations describing  the dynamics of  the perturbations
\begin{mathletters}
\label{pgslin}
\begin{eqnarray}
\label{pgllin}
\partial_t p_g^{(1)} &=& 
- \Big( f_{cat}^{(0)} + v_g \partial_l \Big) p_g^{(1)} 
-  p_g^{(0)} f_{cat}^{(1)}\,,
 \\
\label{psllin}
\partial_t p_s^{(1)} &=& v_s \partial_l  p_s^{(1)}  
       + f_{cat}^{(0)}\, p_g^{(1)} + p_g^{(0)} f_{cat}^{(1)}\,,  \\
\label{ctnl}
\partial_t c_t^{(1)} &=& - \, \alpha c_t^{(1)} \nonumber \\
                    && -\gamma \int^{\infty}_0 d{\it l}  \,\, 
           \Big( v_g p_g^{(1)}
+ \alpha\, {\it l} (  p_g^{(1)} +  p_s^{(1)}) \Big) \,.
\end{eqnarray}
\end{mathletters}
Here $f_{cat}^{(1)}$ is the first order contribution of
an expansion of the  
catastrophe rate $f_{cat}=f_{cat}^{(0)}+f_{cat}^{(1)}+ \ldots$
 with respect to the perturbation 
 $\ct{1}$:
\begin{equation}
\label{ctlin}
f_{cat}^{(1)} = - \fcat \, \frac{\ct{1}}{c_f} \, .
\end{equation}
Since the first order linear equations  (\ref{pgslin})
have  constant coefficients, their solutions depend
exponentially in time  and  $c_t^{(1)}$ 
may be written as
\begin{equation}
\label{A_1}
c_t^{(1)} = A \, e^{\sigma t} + c.c. \,  \,
\end{equation}
(c.c.=conjugate complex). With this ansatz the three equations 
in (\ref{pgslin})  can easily be integrated 
and the solutions  
for the growing and shrinking microtubules are given by
\begin{mathletters}
 \label{pgslin0}
\begin{eqnarray}
\label{pglin}
p_{g}^{(1)} &=& -\frac{\nu f_{cat}^{(0)}  }{v_{g} c_{f}\sigma} 
 \exp{\left(\sigma t -\frac{f_{cat}^{(0)}}{v_{g}} {\it l} \right)}\, \nonumber\\ 
& &\Big[ \exp{\left(-\frac{\sigma}{v_{g}}{\it l}\right )}  
- 1 \Big] \, A    \; + \;  c.c. \, \,,\\
\label{pslin}
p_{s}^{(1)} &=& -\frac{\nu f_{cat}^{(0)}  }{v_{s} c_{f}\sigma}\, 
 \exp{\left(\sigma t -\frac{f_{cat}^{(0)}}{v_{g}} {\it l} \right)}
\Big[ k_{1}
\exp{\left(-\frac{\sigma}{v_{g}} {\it l} \right)} \nonumber\\  & +& k_{2}   
 + K\cdot \exp{\left(\frac{\sigma}{v_{s}}{\it l}\right)} \Big]\, A \; +\; c.c. \, \,.
\end{eqnarray}
\end{mathletters}
Herein we have introduced the   abbreviations 
\begin{eqnarray}
k_{1} &=& 
\frac{f_{cat}^{(0)} }{ 
 \sigma(1 +\beta ) + f_{cat}^{(0)}} \,, \nonumber\\
k_{2} &=& \frac{ \sigma -f_{cat}^{(0)}}
{  f_{cat}^{(0)} + \sigma \beta } \, ,
\end{eqnarray}
and the  boundary condition in Eq.~(\ref{boundps}) requires a vanishing 
integration constant  $K=0$.  
The boundary condition for the time--dependent part of 
the growing microtubules, $p_g^{(1)}(l=0,t)=0$, is also fulfilled.
According to the analytic expressions  for $p_g^{(1)}$ and   $p_s^{(1)}$ given
in Eqs.~(\ref{pgslin0})   both 
  may be 
eliminated in    Eq.~(\ref{ctnl}). The remaining integral in  Eq.~(\ref{ctnl})
can  be calculated analytically  
and the  nonlinear dispersion relation for $\sigma$\, follows
\begin{eqnarray}
\label{dispers1}
&&1 +  \sigma\,  (\sigma + \alpha)\,G \, + \frac{\alpha}{ f_{cat}^{(0)}} \left(
1 +  \beta\, \frac{ f_{cat}^{(0)} -\sigma }{  f_{cat}^{(0)} + \sigma \beta  } 
\right)       \,   \\  
&& - \frac{ f_{cat}^{(0)} } { f_{cat}^{(0)}+ \sigma}\left[
1 + \frac{ \alpha} { f_{cat}^{(0)}+ \sigma} \left( 1 + 
\beta \,\frac{ f_{cat}^{(0)}} 
{ f_{cat}^{(0)} + \sigma ( 1+ \beta )  } \right) \right] =0 \nonumber \,,
\end{eqnarray}
with a reduced  parameter 
\begin{equation}
\label{Gexp} 
G =  \,\, \frac{c_f}{\gamma\nu v_g}
\end{equation}
 for the catastrophe rate
given in Eq.~(\ref{rateexp}) and  with 
\begin{equation}
\label{Glin}
G =  \,\, \frac{f^{(0)}_{cat}}{\bar f \gamma\nu v_g}\,
\end{equation}
for the rate given in Eq.~(\ref{ratelin}).
After a  few 
rearrangements  the dispersion relation in Eq.~(\ref{dispers1}) can
be written as  a fourth order polynomial in $\sigma$
\begin{eqnarray}
\label{dispers2}
&&\sigma^4 \, G \beta \left(1+\beta \right) 
+ \sigma^3 \, G\left[\alpha \beta(1+\beta) +
f_{cat}^{(0)}\left( 1 + 3 \beta + \beta^2 \right)\right] \nonumber \\
&&+ \sigma^2 \left[ G \alpha f_{cat}^{(0)} \left(1 + 3 \beta + \beta^2\right) 
+ \left( \beta + 2 G {f_{cat}^{(0)}}^2 \right) \left(1 + \beta \right) \right] \nonumber \\
&&+ \sigma  \left[ \alpha \left(1+\beta\right)\left(1 +\beta + 2G  f_{cat}^{(0)^2}
\,\right) + Gf_{cat}^{(0)^3} \right.  \\
&&+ \left. f_{cat}^{(0)}\left(1 + 2 \beta \right) \right] 
+  \alpha f_{cat}^{(0)} \left[ 2 + 2 \beta + G f_{cat}^{(0)^2} \,\right] +
f_{cat}^{(0)^2} = 0 \nonumber \,.
\end{eqnarray}
This polynomial
describes the linear 
stability  of the stationary solutions
given by  Eq.~(\ref{pgsstat}) and Eq.~(\ref{cgleich0})
 completely and  they are  unstable 
in the parameter range where the {\it growth rate} becomes positive, 
$Re(\sigma)>0$. Keeping for instance all  parameters
besides the dimer concentration $c_0$ fixed,  then 
the {\it neutral stability condition}
$Re(\sigma)=0$  provides 
an equation  for the critical dimer  concentration
 $c_{0c}$.  For concentrations larger than this critical 
value,  $c_0 > c_{0c}$,
 the stationary solutions are  unstable. 

The smallest  critical dimer concentrations $c_{0c}$  
for an oscillatory polymerization 
are required if the parameters  $\alpha$,$\beta$  and $G$
take intermediate values, as discussed in more detail below. 
At the threshold the real part 
  $Re(\sigma)=0$ vanishes  and 
the imaginary part of   $\sigma$  is 
the so-called    Hopf--frequency 
$\omega_c = Im(\sigma)$.   In this special case with a 
purely imaginary  $\sigma$ the 
 polynomial in Eq.~(\ref{dispers2}) can be 
decomposed into its  
real and imaginary parts giving  
two coupled equations for the determination of the
 two unknowns  $f_{cat}^{(0)}$ and $\omega_c$. 
Having determined $f_{cat}^{(0)}$ numerically, $c_{0c}$ may  be calculated 
via Eq.~(\ref{cgleich0}). 

%%%%%%%%%%%%%%%%%%%%%%%%%%%%%%%%%%%%%%%%%%%%%%%%%%%%%%%%%%%%%%%%%%%%%
\subsubsection{Limiting cases with the rate in Eq.~(\ref{rateexp})}
\label{limit1}
%%%%%%%%%%%%%%%%%%%%%%%%%%%%%%%%%%%%%%%%%%%%%%%%%%%%%%%%%%%%%%%%%%%%%

For the  limiting cases  
$\beta \to 0$, $\beta \to \infty$,
$\alpha \to 0$ and  $\alpha \to \infty$ 
 analytical expressions can be given for both  the 
threshold concentration  $c_{0_c}$ 
and the Hopf  frequency $\omega_c$. This is explained at first  
   for the
catastrophe rate  given by Eq.~(\ref{rateexp}) and 
for the parameter $G$ given in  Eq.~(\ref{Gexp}). 
At threshold one  has   $\sigma = i \omega_c$  and two equations follow 
 from the nonlinear dispersion relation in Eq.~(\ref{dispers2})  which 
determine the two unknowns
$\omega_c$ and $f_{cat}^{(0)}$. 
The  critical initial concentration $c_{0c}$ 
follows via  $f_{cat}^{(0)}$ from
Eq.~(\ref{cgleich0}).

\paragraph{ $\alpha \to \infty$:}
In this limit one obtains from (\ref{dispers2}) 
\begin{mathletters}
\begin{eqnarray}
\label{scale1alph}
f_{cat}^{(0)} &=& \frac{1}{\alpha G} \,\, \frac{1 + \beta} {1+\beta+\beta^2}\,, \\
\omega_c &=& \sqrt{ \frac{1+\beta}{G \beta}\,\,}\,.
\end{eqnarray}
\end{mathletters}
Accordingly the critical tubulin concentration diverges as 
 $c_{0c} \propto \alpha^2$, that  agrees with the 
full numerical results shown in Fig.~\ref{fig1}, besides 
small logarithmic corrections. 
In this limit the  Hopf  frequency  $\omega_c$ becomes  independent of
$\alpha$ and   with increasing values of $\beta$ it decreases slightly 
to  a constant value $\omega_c \sim \sqrt{1/G}$. 

\paragraph{$\alpha \to 0$:} In this case $\omega_c \sim \sqrt{1/G}$ becomes 
also independent of $\alpha$  and the catastrophe rate vanishes as $f_{cat}^{(0)} \sim \alpha$.
Therefore the critical tubulin concentration
diverges according to Eq.~(\ref{cgleich0}) as $c_{0c}\sim \alpha^{-2}$.

\paragraph{$\beta \to 0$:} In this limit one obtains 

\begin{equation}
f_{cat}^{(0)} = \sqrt{ \frac{\beta}{G} } \quad {\rm and} \quad
\omega_c = {\left(\frac{\alpha^2}{G}\right)}^{1/4} \,\left(\frac{1}{\beta}\right)^{1/4}\,.
\end{equation}
The Hopf  frequency diverges with increasing values of the
shrinking velocity as $\omega_c \sim v_s^{1/4}$ in agreement with
the numerical results shown in  Fig. \ref{fig2}.
With this expression for   $f_{cat}^{(0)}$ one obtains via
Eq.~(\ref{cgleich0}) for the critical initial
concentration  $c_{0c} \sim  G/ \beta + c_f \ln{( f (G/\beta)^{1/2})}$.
For medium parameter values this is essentially 
$c_{0c} \propto 1/\beta$ as  indicated in Fig. \ref{fig2}.

In experiments the shrinking velocity was always larger then
the growth velocity, therefore the limit $\beta  \to \infty$ is discarded.

\subsubsection{Limiting cases for the rate in Eq.~(\ref{ratelin})}

The tendencies for the parameter dependence of the threshold
for the Hopf   bifurcation as discussed in 
Sec.~\ref{limit1}
are by far not a special property of the
choice of the catastrophe rate in Eq. (\ref{rateexp}).
Therefore we consider the same limiting cases as before for
the catastrophe rate given in Eq.~(\ref{ratelin}) and with $G$ as 
defined in Eq. (\ref{Glin}).

{\it a. $\alpha \to \infty $:}
In this limit one obtains from Eq. (\ref{dispers2})

\begin{mathletters}
\begin{eqnarray}
f_{cat}^{(0)} &=& \left( \frac{g(1+\beta)}{\alpha(1+\beta+\beta^2)} \right )^{1/2} \, ,\\
\omega_c &=& \frac{ \left( \alpha g (1+\beta+\beta^2) (1+\beta)\right )^{1/4}}{\beta^{1/2}} \, ,
\end{eqnarray}
\end{mathletters}
with $g= \bar f \gamma \nu v_g$.
Hence the critical tubulin  concentration required for a
Hopf bifurcation diverges as  $c_{0c} \propto \alpha$.

{\it b. $\alpha \to 0$:} In this limit one has $f_{cat}^{(0)} \propto \alpha$ 
and $c_{0c} \propto \alpha^{-2}$ diverges too.

{\it c. $\beta \to 0$:}
For this limit we obtain 
\begin{mathletters}
\begin{eqnarray}
f_{cat}^{(0)} &=& {(g \beta)}^{1/3}   \, ,\\
\omega_c &=&  g^{1/6} \alpha^{1/2} \left(\frac{1}{\beta}\right)^{1/3}  \, .
\end{eqnarray}
\end{mathletters}      
This confirms the importance of a finite ratio 
of $\beta = v_g/v_s$, because the threshold diverges for $\beta \to 0$,
similar as for the catastrophe rate given in Eq.~(\ref{rateexp}). 

%-----------------------------------------------
\subsubsection{Traveling waves solutions}
\label{LintravelI}
%-----------------------------------------------

At the  threshold of the Hopf bifurcation the rate $\sigma$ is
purely imaginary,
$\sigma = i \omega_c$, and  
 the expressions  given in Eq.~(\ref{A_1}) and Eqs.~(\ref{pgslin0}) 
are  oscillatory in time
\begin{mathletters}
\label{eq:lin}
\begin{eqnarray}
\label{eq:linc}
c_t^{(1)} &=& 2 \, A\, \cos(\omega_c t) \,, \\
\label{eq:linpg}
p_{g}^{(1)} &=&  \frac{S_1}{v_g}
\exp(-\frac{f_{cat}^{(0)}}{v_{g}} {\it l} )
\Big[ \sin (\omega_c t) -  \sin (\omega_c (t- {l}/{v_g})) \Big] \,, \\
\label{eq:linps}
p_{s}^{(1)} &=& - \frac{S_1}{v_s}
\exp(-\frac{f_{cat}^{(0)}}{v_{g}} {\it l} ) \left [\, k_2 \, \sin (\omega_c t 
+ \varphi_2 ) \right. \nonumber \\
&+& \left. k_1 \, \sin (\omega_c (t- {l}/{v_g}) +\varphi_1 ) \,\right ] \, \,,
\end{eqnarray}
\end{mathletters}
whereby  the following  abbreviations for the amplitudes 
\begin{mathletters}
\begin{eqnarray}
S_1 &=&    \frac{ 2 A \nu f_{cat}^{(0)}  } {\omega_{c} \,c_{f}} \,, \\
k_1 &=&  \frac{ \sqrt { {\fcat}^4 + \omega_c^2 {\fcat}^2
 (1+\beta)^2}} { {\fcat}^2+\omega_c^2(1+\beta)^2 } \,, \\
 k_2 &=& \frac{ \sqrt {\left(\omega_c^2 \beta - {\fcat}^2 \right)^2 + {\fcat}^2 \omega_c^2
 \left(1+\beta \right)^2 }} {{\fcat}^2+\left( \omega_c \beta \right)^2} \ ,
\end{eqnarray}
\end{mathletters}
and phases
\begin{mathletters}
\begin{eqnarray}
  \label{eq:linabb}
\varphi_1 &=& - \arctan \left( \frac{\omega_c(1+\beta) }{f_{cat}^{(0)}  }\right) \,, \\
\varphi_2 &=&  \arctan \left( \frac{\omega_c  f_{cat}^{(0)} (1+\beta) }{ \beta 
\omega_c^2 - f_{cat}^{(0)^2}  }    \right ) \, ,
\end{eqnarray}
\end{mathletters}
have been introduced. 
The analytical expressions 
for $p_g^{(1)}$ and $p_s^{(1)}$ indicate that the 
time--dependent contribution to the length distribution of 
the microtubules  
includes homogeneous amplitude oscillations 
and  waves
with a  wavelength  $\omega_c/v_g$
 that travel to larger values
of the length $l$.
 Hence, the length 
distributions $p_{g,s}^{(1)}$
depend on two different length scales, the decay length
$v_g/f_{cat}^{(0)}$ and the wave length $v_g/\omega_c$ of
the traveling waves. 
With the explicit solutions for
$c_t^{(1)}$ and $p_g^{(1)}$ the phase of the 
oscillating part  of the tubulin--d concentration
relative to the phase of $c_t^{(1)} $ as well as its 
oscillation amplitude is 
calculated via  Eq.~(\ref{ctcd1}).

%%%%%%%%%%%%%%%%%%%%%%%%%%%%%%%%%%%%%%%%%%%%%%%%%%%%%
\subsubsection{Numerical results for the  threshold of model I }
\label{resthreshI}
%%%%%%%%%%%%%%%%%%%%%%%%%%%%%%%%%%%%%%%%%%%%%%%%%%%%%

Since $\sigma$ is purely imaginary at the threshold of a Hopf bifurcation, 
 $\sigma = i \omega_c$, the  dispersion 
relation in Eq.~(\ref{dispers2}) can be decomposed in its
real and imaginary part and from these  two equations
the   critical concentration $c_{0c}$ and the frequency $\omega_c$ 
may be calculated numerically  as a function 
 of  the  parameters.  The numerical 
calculations in this section are restricted  to the
exponential tubulin dependence of the
  catastrophe rate as given by Eq.~(\ref{rateexp}).

\begin {figure}
\epsfxsize 8.7cm \ifpreprintsty \epsfxsize 15.0cm \fi
\epsfbox {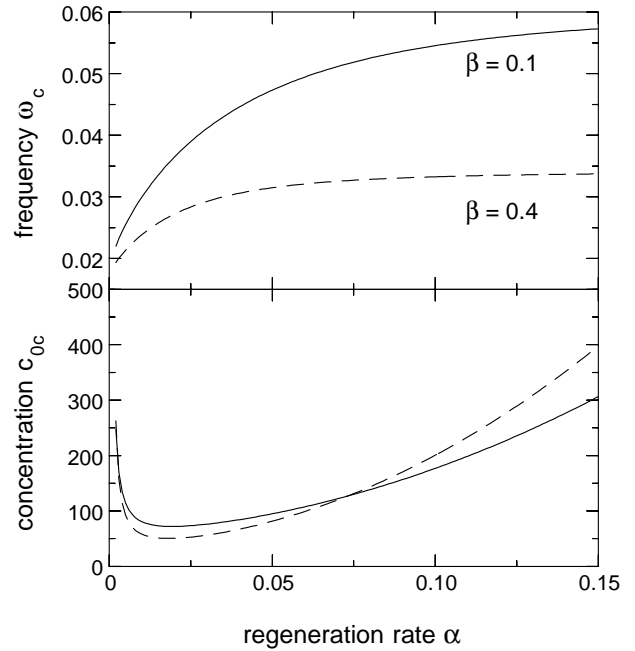}
\caption [Figure 3] 
  {The critical tubulin dimer  concentration  $c_{0c}$ 
and the critical oscillation frequency 
$\omega_c$   are given 
at the threshold of  the Hopf bifurcation as a function of
the regeneration rate $\alpha$,  for two values 
of $\beta = v_g/v_s$ and for the   
constant $G=3000$.
The catastrophe rate given in Eq.~(\ref{rateexp}) has been used
with the parameter values $f=0.1$ and $c_f=3$. 
}
\label{fig1}
\end{figure}

\begin{figure} 
\epsfxsize 8.7cm \ifpreprintsty \epsfxsize 15.0cm \fi
\epsfbox {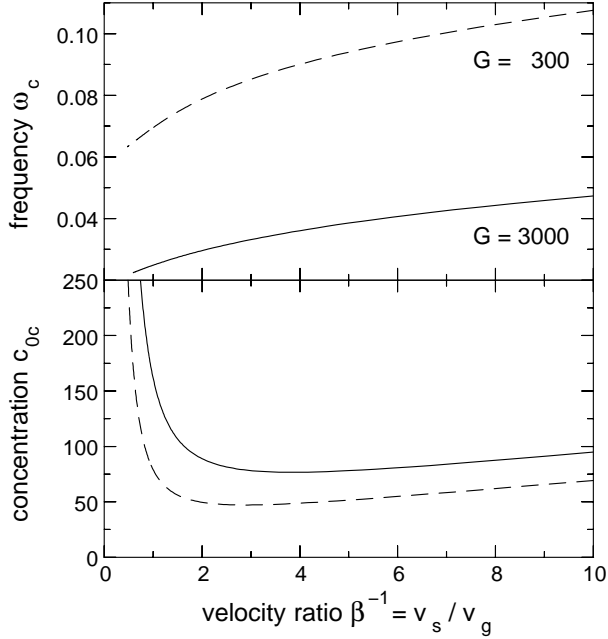}
\caption [Figure 3] 
  {The critical tubulin  concentration 
$c_{0c}$  and the  frequency $\omega_c$ are given at the
Hopf  bifurcation as a function 
of the ratio between the shrinking and growth velocity 
of microtubules,  $v_s / v_g = \beta^{-1}$ , for 
$\alpha=0.05$ and  for  two  
different values of  $G$. The $c_t$--dependence of the 
catastrophe rate as given in  Eq.~(\ref{rateexp}) has been used
with the same  parameters as in Fig.~\ref{fig1}.
}
\label{fig2}
\end{figure}

The critical tubulin dimer concentration  
$c_{0c}$ and the critical 
frequency $\omega_c$ at the Hopf bifurcation
 are shown in Fig.~\ref{fig1}
as  function of the regeneration rate $\alpha$
and in  Fig.~\ref{fig2} as function of 
the velocity ratio $1/\beta=v_s/v_g$, whereby
the reduced  parameter $G$  has been chosen 
at the values $G=3000$ and $G=300$, respectively.
Since $G$  includes a number of parameters
 the  curves in both figures  represent a larger
parameter set.  In the limit
of a vanishing regeneration
 and in the limit of very large values of 
$\alpha$, where the
regeneration process is  much faster than any 
other process, the critical tubulin concentration $c_{0c}$ 
diverges and  therefore the Hopf bifurcation is suppressed. 
In addition, both figures indicate 
that the  smallest  values of the critical tubulin concentration $c_{0c}$ 
are  obtained at  intermediate values 
of the  parameters $\alpha$, $\beta$ and $G$. The location of
the threshold minima, however,  depends on the actual values of the rest of the
 parameters. 
The frequency $\omega_c$ becomes rather small  in the limit $\alpha  \to 0$ and 
for large values of   $\alpha$ this frequency 
becomes   independent of it, cf. section \ref{limit1}.

In Fig.~\ref{fig2}
the threshold minimum 
is less pronounced than in Fig.~\ref{fig1} and  in 
the limit 
$\beta = v_g/v_s \to 0$ the 
threshold $c_{0c}$ increases linearly with $v_s$ and in agreement
with limits given in section \ref{limit1}.
Accordingly, there  is no Hopf bifurcation 
for the  reduced model that follows in the  limit 
$\beta  \to 0$ 
as described in  section \ref{modred}. Hence, the 
dynamics of shrinking   microtubules  is  one  essential degree of
freedom favoring  oscillating microtubule polymerization.
The dynamics of oligomers, as discussed in 
Sec.~\ref{osthreshII}, is an alternative
degree of freedom that favors oscillations.

% ******************************************************************************
%   Figure 3
% ******************************************************************************
%
%****************************************************************************

For large values of $v_s$  the frequency $\omega_c$ becomes 
large too 
and the oscillation period    becomes  much shorter  than any   relaxational 
dynamics of
$p_g$ and $c_t$. According to the 
quick shrinking, the life time of a depolymerizing  microtubules 
vanishes and therefore the amplitude of the density of shrinking
microtubules is  small too, $p_s \propto 1/v_s$.
In other words, in the limit of large values of $v_s$
the intermediate step of shrinking microtubules may
be neglected and the transition
from  $p_g$ to tubulin-d dimers is effectively  a direct 
process as explicitly assumed for the reduced model.
If either the regeneration or the shrinking dynamics 
becomes too fast,  the  Hopf bifurcation is suppressed. 
The two  intermediate steps, the   depolymerization  and 
the regeneration,  
act obviously as antagonistic steps or jam 
processes that favors oscillations.

Since  one has at threshold $\sigma = i\omega_c$, 
$p_g^{(1)}$ and $p_s^{(1)}$
in Eq.~(\ref{eq:linpg}) and Eq.~(\ref{eq:linps}) 
include both  traveling wave contributions  
$\propto \, \exp[-i(\omega_c t - k l)]$ with a wave number $k=\omega_c/v_g$
and that  always  travel towards 
larger lengths of the microtubules.    
The length distribution 
is exponentially decaying on 
the  length scale  $v_g / \fcat$. If this  is large 
compared to the wave length 
$\lambda = 2 \pi v_g/\omega_c$, as  for instance in the limit 
$v_s \gg v_g$, then one has a kind of self averaging in the respective 
integrals 
and the Hopf bifurcation is suppressed.

% ******************************************************************************
%   Figure 4
% ******************************************************************************
%
%*************************************************************************

\begin{figure}
\epsfxsize 8.7cm \ifpreprintsty \epsfxsize 15.0cm \fi
\epsfbox {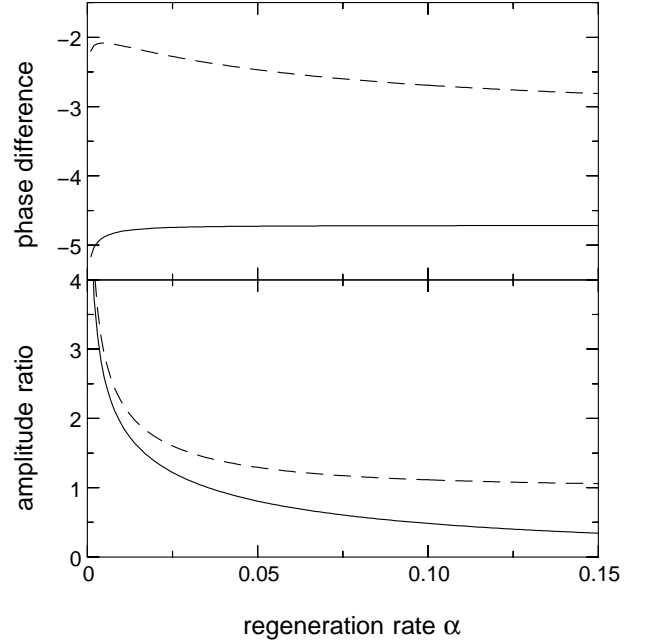}
\vspace{2mm}
\caption %[Figure 3] 
  {In the upper part the differences between the 
phases of  the oscillating contributions  of the tubulin-d concentration  $c_d$ (solid) 
and of the tubulin-t $c_t$ as well as between  
 $L(t)$ and $c_t$ 
(dashed)   are shown as function of the regeneration rate 
$\alpha$. 
In the lower part the ratios  between the amplitudes of 
the oscillating contributions $c_d^{(1)}$ and $c_t^{(1)}$
(solid) 
as well as between the amplitudes of $L^{(1)}$ and $c_t^{(1)}$ (dashed)
 are shown.
The rest of parameters are as in Fig.~\ref{fig1}.
}
\label{figphaseps}
\end{figure}

The phase difference  between the oscillations of
tubulin--t and the oscillations of the 
total amount of polymerized tubulin, 
described by $L(t)$  in  Eq.~(\ref{conserv}), 
is another experimentally accessible   quantity  
\cite{Mandelkow:94.3}.  The  difference 
between the phases  of the oscillatory contributions 
of $c_d$ and $c_t$ as well as  the difference between the phases of 
$L(t)$ and $c_t$ 
are given in   Fig.~\ref{figphaseps}. 
These phase differences as well as the ratios between the 
amplitudes of the fields, cf. lower part of Fig.~\ref{figphaseps},
are calculated at the threshold of the Hopf  bifurcation 
by using 
the analytical solutions calculated 
in Sec.~\ref{LintravelI}.

For large values of $\alpha$,  tubulin--d 
is quickly regenerated
into tubulin--t  and therefore the
density $c_d$ becomes  smaller as shown by the lower
part in  Fig.~\ref{figphaseps}. 
In the opposite limit of small values of $\alpha$
tubulin--t  is consumed by nucleation and growth 
of microtubules, but the  source, which is supplied by  
the regeneration of tubulin--d, decays and therefore one obtains large
values for the 
 ratio between the amplitude of $c_d^{(1)}$
and   $c_t^{(1)}$ as well as between the amplitudes of 
$L^{(1)}$ and $c_t^{(1)}$.

The decay of the ratio between the amplitudes
of  $L^{(1)} $ and $c_t^{(1)} $ is less obvious.
$L^{(1)}  = L^{(1)} _g+L^{(1)} _s$ has the two contributions 
$L^{(1)} _g =\gamma \int_0^{\infty} dl \,l \,p_g^{(1)} = {\bar A} \cos(\omega_c t+ \varphi_{\bar A})$
and 
$L^{(1)} _s =\gamma \int_0^{\infty} dl \,l \,p_s^{(1)} = {\bar B} \cos(\omega_c t+ \varphi_{\bar B})$.
The two  amplitudes ${\bar A}$ and ${\bar B}$ increase with the regeneration rate $\alpha$.
However, with increasing values of $\alpha$ 
the  phase difference $\varphi_{\bar A} -\varphi_{\bar B}$ increases as well up to unity 
 leading  to an effective  decay of the sum 
 $L^{(1)} $ as shown 
by the lower part of  Fig.~\ref{figphaseps}.

% ******************************************************************************
%   Figure 5
% ******************************************************************************
%
%*********************************************************************************

The phase  shifts of $c_d$ and $L(t)$ 
with respect to the phase of $c_t$ 
are rather independent of the regeneration rate $\alpha$
as  shown in 
Fig.~\ref{figphaseps}. The absolute values of these shifts 
are in  qualitative agreement with the expectation as described in 
the following.
 At the  maximum of $c_t$ the catastrophe rate
takes its minimum and therefore, since the nucleation and the growth 
velocity are constant,  
$L(t)$ is increasing for a while and up to the moment when
enough $c_t$ is consumed and the catastrophe rate 
increases again.  
Due to an increasing  decay of microtubules  the maximum of the latter will also
lead to a delayed maximum for $c_d$. 
For large values of $\alpha$ the amount of polymerized 
tubulin is nearly in anti phase with respect to $c_t$, which is also
mentioned in Ref.~\cite{Mandelkow:94.1}. 
The slightly stronger $\alpha$--dependence of the phase
difference between  $L(t)$ and $c_t$  is mainly due to the
$\alpha$-dependence of shrinking microtubules $p_s$, because 
the relative phase of $p_g^{(1)}$ is nearly independent 
of $\alpha$  (see also Section \ref{osthreshII}).

%---------------------------------------------
\subsection{Model II}
\label{osthreshII}
%---------------------------------------------

Here the stability of the stationary
polymerization state of model II, 
described by 
$c_{t}^{(0)}$,$c_{d}^{(0)}$ and $p_g^{(0)}$,
 is investigated with
respect to small perturbations $c_{t}^{(1)}$, $c_{d}^{(1)}$  and  $p_g^{(1)}$. 
With the ansatz  
\begin{mathletters}
\begin{eqnarray}
  p_{g} &=&\p{g}{0}+ \p{g}{1} \, , \\
        c_{t,d} &=& c_{t,d}^{(0)} + c_{t,d}^{(1)}  \,\, ,
\end{eqnarray}
\end{mathletters}
the equations for model II  are linearized with 
respect to these perturbations and one obtains 
the following set of linear 
 differential equations with constant coefficients
\begin{mathletters}
\label{olict}
\begin{eqnarray}
  \partial_t \p{g}{1}&=& -  f_{cat}^{(1)} \, \p{g}{0} 
                          -\left( \fcat + v_g \partial_l \right ) \p{g}{1} \, ,\\
\label{olict1}
  \partial_t \ct{1}  &=& -\eta \lambda v_g \int_0^{\infty} dl \,  \p{g}{1} 
                                          + \alpha \cd{1} \, , \\
\label{olicd1}
  \partial_t \cd{1}  &=& -\chi \left( \ct{1} + \cd{1} 
                                   + \eta \lambda \int_0^{\infty} dl \, l\, \p{g}{1}  \right)
                                - \alpha \cd{1} \, .
\end{eqnarray}
\end{mathletters}
$f_{cat}^{(1)}$ is the first order correction 
with respect to its value in the stationary state 
and it is given  by     equation (\ref{ctlin}).

The time--dependent contributions to the  tubulin--t and tubulin--d dimer densities  are
described by 
\begin{mathletters}
\label{campliII}
\begin{eqnarray}
 \ct{1} &=& A \, \, e^{\sigma t} + c.c. \,\, ,  \\
 \cd{1} &=& E \,A \, e^{\sigma t} + c.c. \,\, ,
\end{eqnarray}
\end{mathletters}
with the common complex amplitude $A$ and 
relative complex factor $E$ that describes
via  $E= |E| e^{i\varphi_d}$ the amplitude ratio $|E|$
 as well as the phase difference $\varphi_d$ 
between  both fields.
With  the solution for
growing microtubules as given in 
Eq.~(\ref{pglin}) we can   eliminate  $c_t^{(1)}$ and $p_g^{(1)}$ 
in  Eq.~(\ref{olict1}) and 
we again obtain  from  the resulting
solubility condition a
nonlinear dispersion relation for the exponential factor $\sigma$

\begin{eqnarray}
 \label{disoli1}
  &&\left( G \sigma + \frac{1}{\sigma+\fcat}\right) ( \sigma+\alpha+\chi) \nonumber \\
  &+& \alpha \chi
   \left( G + \frac{\sigma+2\fcat}{\fcat\left(\sigma+\fcat\right)^2} \right ) =0 \, ,
\end{eqnarray}
with $G=c_f/(\eta \lambda v_g \nu)$.
After a few rearrangements of this equation
one obtains a 
fourth order polynomial in $\sigma$ for model II as well
\begin{eqnarray}
\label{disoli2}
  &&\sigma^4 \fcat \,G + \sigma^3 \fcat \,G \left( 2\fcat 
   + \alpha+\chi \right) \nonumber \\
  &+& \sigma^2 \fcat \left[ 1+\alpha\chi G + G \fcat \left( 2\chi+\fcat 
          +2\alpha \right) \right ] \nonumber \\
 &+&  \sigma \left[ \fcat \left( \fcat+\alpha+\chi \right ) 
  + \alpha\chi \right. \nonumber \\ 
 &+& \left.  {\fcat}^2 G    \left( 2\alpha\chi  
  + \fcat(\alpha+\chi) \right) \right ]  \nonumber \\
 &+& \fcat \, \alpha \chi \left( G {\fcat}^2 +2 \right) 
  + {\fcat}^2 ( \alpha + \chi ) = 0  \,\, ,
\end{eqnarray}
that determines the linear stability of the stationary 
polymerization for model II.   
Again we 
 are interested in the neutrally stable case, $Re(\sigma)=0$,  that
separates 
the stable  from the unstable regime.
At the neutral stability point of the Hopf bifurcation  one has 
  $\omega_c = Im(\sigma)$ and  
Eq.~(\ref{disoli2}) can be decomposed into  its
real and imaginary part.
From these two equations $\fcat$ and $\omega_c$
are determined by standard methods.
 $c_{0c}$
may  be calculated via Eq.~(\ref{statoli}).

%-------------------------------------------
\subsubsection{Traveling waves solutions}
\label{Lintravell_II}
%-------------------------------------------

At the Hopf  bifurcation 
the  non-stationary part of  growing microtubules
is again described by the distribution    given by Eq. (\ref{eq:linpg}) 
and the fields  $c_{oli}$ and $c_d$ are not in phase
with  $c_t$ in general. The two fields may
be written in terms of 
the amplitude ratio $|E|$ and the relative phase $\varphi_{d}$
 in the following form 
\begin{mathletters}
\label{ollin}
\begin{eqnarray}
  \ct{1} &=& 2 \, A \, \cos( \omega_c t) \, ,\\
  \cd{1}&=& 2 \,A \,|E| \cos( \omega_c t + \varphi_d)  \, .
\end{eqnarray}
\end{mathletters}
The amplitude ratio $|E|$ and the phase 
$\varphi_d$ can  be determined from the two 
 coupled equations  (\ref{olict1}) and (\ref{olicd1})
and they are given by 
\begin{eqnarray}
  |E| &=& \frac {\sqrt{\, {\fcat}^2 + \omega_c^2 \,( G( {\fcat}^2+\omega_c^2)-1)^2 } }
             { \alpha G ( {\fcat}^2 + \omega_c^2 )} \, , \nonumber \\
  \varphi_d&=& \arctan \left( \frac{\omega_c}{\fcat}  
\left(G({\fcat}^2+\omega_c^2) -1\right)
                   \right ) \, .
\end{eqnarray}

In a similar manner the oligomer density 
$c_{oli}^{(1)}$ may be written in terms of an 
amplitude ratio $|F|$  and
a relative phase $\varphi_{oli}$   between
 $c_{oli}^{(1)}$ and $c_t^{(1)}$
\label{ollinoli}
\begin{eqnarray}
  c_{oli}^{(1)}&=& 2 \, A\, |F| \cos( \omega_c t + \varphi_{oli})  \, ,
\end{eqnarray}
with 
\begin{eqnarray}
  |F| &=& \frac { \sqrt { \alpha^2 +\omega_c^2}} {\chi} \,\,|E|\, ,
\qquad {\rm and}  \nonumber \\
  \varphi_{oli} &=& \arctan \left(\frac{\alpha \tan(\varphi_d) + \omega_c }
{\alpha - \omega_c \tan(\varphi_d) } \right)\, .
\end{eqnarray}

The oscillatory contribution to the
polymerized tubulin 
 $L^{(1)}$ can also be written as 
a harmonic function, $L^{(1)}=2A |H| \cos{(\omega_c t + \varphi_L)}$.
For both,  the  amplitude ratio $|H|$ and the relative 
phase $\varphi_L$,  one obtains 
long expressions that are not presented here.
The phase shifts and the amplitude ratios  between the
oscillating fields  are  shown in Figure \ref{figphaseoli}
 as function of the regeneration rate $\alpha$. As discussed in
Section \ref{osthreshI}, the phase shift of the  polymerized tubulin 
$L^{(1)}(t)$ with respect to $\ct{1}$ is rather independent
of  $\alpha$, whereas the phase of oligomer oscillations change slightly with $\alpha$. 
A phase shift $\pi$ between the polymerized tubulin  and oligomers
is measured in experiments, cf.  
Refs. \cite{Mandelkow:92.1} and \cite{Mandelkow:90.3}. 
In this model this
is only possible in the limit of a dissociation rate $\chi$
much smaller than the regeneration rate $\alpha$.

%%%%%%%%%%%%%%%%%%%%%%%%%%%%%%%%%%%%%%%%%%%%%%%%%%%%%
\subsubsection{Numerical results for the  threshold of model II}
\label{resthreshII}
%%%%%%%%%%%%%%%%%%%%%%%%%%%%%%%%%%%%%%%%%%%%%%%%%%%%%

At the threshold one has again 
 $\sigma = i \omega_c$ and  from the  imaginary together with  the real
part of the dispersion 
relation in Eq.~(\ref{disoli2}) the  
  critical concentration $c_{0c}$ and the Hopf frequency $\omega_c$ 
may be calculated as  function 
 of  the  parameters. Also for model II  we restrict
 our numerical calculations 
to the catastrophe rate with the exponential dependence 
given in Eq.~(\ref{rateexp}).

% ******************************************************************************
%   Figure 6
% ******************************************************************************
%
\begin{figure}
\epsfxsize 8.7cm \ifpreprintsty \epsfxsize 15.0cm \fi
\epsfbox {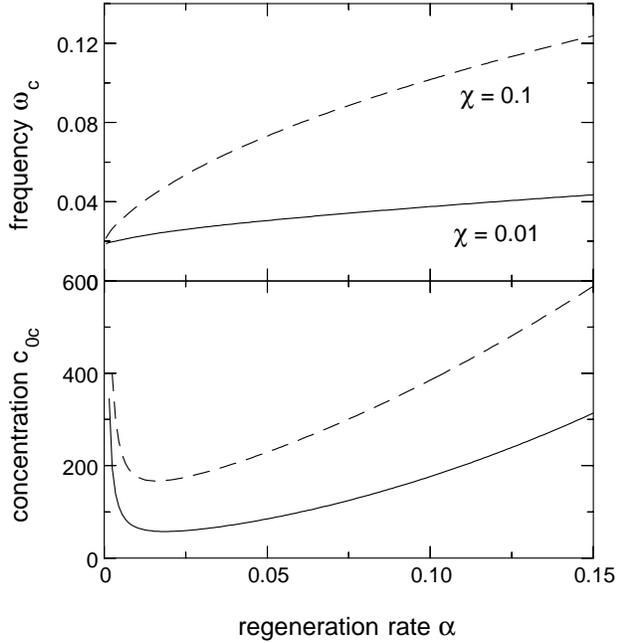}
\caption [Figure 3] 
  {For model II the critical tubulin  concentration $c_{0c}$ 
and the critical  frequency $\omega_c$ are shown  
at the Hopf bifurcation  as a function of
the regeneration rate $\alpha$ and for two different values
of the decay rate $\chi$ of oligomers. 
The rest of the parameters are $G=3000$, $f=0.1$ and
 $c_f=3$.
}
\label{fig3}
\end{figure}
%*******************************************************************

The critical tubulin concentration 
$c_{0c}$ and the critical 
frequency $\omega_c$ at the Hopf bifurcation
 are shown in Fig.~\ref{fig3}
as  function of the regeneration rate $\alpha$ and for two different values of the
decay rate of oligomers $\chi$,  whereby
for the reduced parameter $G$  the value  $G=3000$ has been chosen.
Since $G$  includes a number of parameters
 the  curves in both parts  represent a larger
parameter set.  For a fixed finite value for $\chi$ in the limit 
of a vanishing regeneration   $\alpha \to 0$, where the polymerization cycle 
is interrupted,  
 and in the limit of very large values of 
$\alpha$, where the
regeneration process is  much faster than any 
other process, the critical tubulin concentration $c_{0c}$ 
diverges similar as for model I and therefore the  Hopf bifurcation is   suppressed. 
If $\alpha$ is kept fixed at a medium value the threshold curve $c_{0c}(\chi)$ 
as function of the decay rate $\chi$ for oligomers 
  has a similar shape as shown as function of   $\alpha$
in Fig.~\ref{fig3}.
The critical tubulin concentration $c_{0c}$  also takes its smallest values
at  intermediate values 
of  $\alpha$, $\chi$  and $G$, whereby the location of
the threshold minima depends on the actual values of the rest of parameters. 
With a decreasing rate $\alpha \to 0 $ of tubulin regeneration also 
the frequency  $\omega_c$ becomes  small. On the other
hand for large values of   $\alpha$ the tubulin regeneration 
is not anymore a rate limiting factor and the critical frequency 
$\omega_c$ becomes  rather independent of $\alpha$ , cf. section \ref{limit1}.

% ******************************************************************************
%   Figure 7
% ******************************************************************************
%  
\begin{figure}
\epsfxsize 8.7cm \ifpreprintsty \epsfxsize 15.0cm \fi
\epsfbox {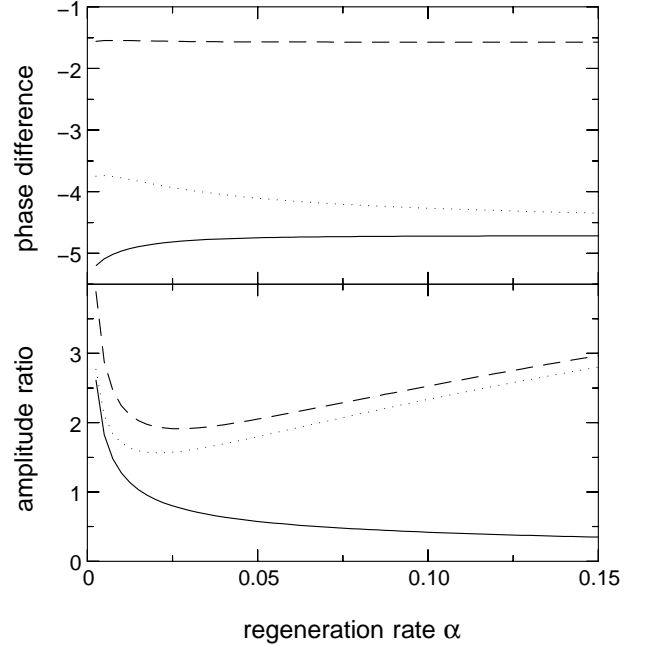}
%
%\vspace{0.3cm}
\caption %[Figure 3] 
  { The phase differences (upper part)  and the amplitude ratios (lower part)
between  the oscillating contributions to 
the tubulin-d concentration $c_d^{(1)}$ (solid), the oligomer concentration
$c_{oli}^{(1)}$ (dotted) and the total polymerized tubulin $L^{(1)}(t)$ (dashed) with
respect to the tubulin-t concentration $c_t^{(1)}$  is shown 
as a function of the regeneration rate $\alpha$.  The other parameters are  $G=3000$ and $\chi=0.02$.
}
\label{figphaseoli}
\end{figure}
%*************************************************************************************

The stability of oligomers and therefore the decay rate $\chi$ depend very much 
on the available GTP: Increasing  GTP concentrations destabilize oligomers and increase
the decay rate $\chi$ \cite{Mandelkow:88.1,Job:89.1}. 
With increasing GTP concentrations also
the  rate $\alpha$ of the transition from  $c_d$ to $c_t$ is enhanced. 
However, if the tubulin regeneration and the oligomer decay become to quick, 
an oscillatory polymerization is suppressed. In other words, if one increases 
$\alpha$ and $\chi$ beyond some 
minimum values, 
the threshold concentration for tubulin increases too.
Such a  tendency for the GTP dependence  of the oscillation is in agreement with the
results  reported from experiments  \cite{Mandelkow:88.1,Job:89.1,Mandelkow:92.1}.

With increasing values of $\alpha$ tubulin--d 
is again quickly transfered by the regeneration process into
tubulin--t, leading to a small amplitude ratio
 $c_d^{(1)}/c_t^{(1)}$. Accordingly more tubulin is left to be stored
in $L^{(1)}$ and $c_{oli}^{(1)}$. Therefore both increase with
larger values of $\alpha$  as indicated in
Fig.~\ref{figphaseoli}. This has to be compared
with $L^{(1)}$ for model I, where it decays as function of $\alpha$
because the phase shift between the contributions
of the growing and shrinking microtubules 
changes too. For model II the
relative phases are also rather independent of $\alpha$, whereby
due to the quick regeneration  of $c_d$ the relative phase 
between $c_t^{(1)}$ and $c_d^{(1)}$ is slightly decreasing.

%------------------------------------
\subsubsection{Reduced models}
%------------------------------------

The dispersion relation for model I and II, 
 considered in the previous section,
became equivalent in the 
limits $\beta \to 0$ and
$\chi \to \infty$ 
and in both cases one obtains 
the same dispersion relation
\begin{eqnarray}
&& \sigma^3 \, G \fcat 
+ \sigma^2 \left[ G \alpha f_{cat}^{(0)}  
+ 2 G {f_{cat}^{(0)}}^2  \right] \nonumber \\
&+& \sigma  \left[ \alpha \left(1 + 2G  f_{cat}^{(0)^2}
\,\right) + Gf_{cat}^{(0)^3}  + f_{cat}^{(0)} \right]  \nonumber \\
&+&  \alpha f_{cat}^{(0)} \left[ 2 + G f_{cat}^{(0)^2} \,\right] +
f_{cat}^{(0)^2} = 0 \, ,
\end{eqnarray}
with the reduced parameter $G$ as given in Eq.~(\ref{Gexp})
for model I and with $G=c_f/(\eta\lambda v_g \nu)$ for model II.
This polynomial in $\sigma$  has always negative
growth rates, $Re(\sigma)<0$, and therefore 
stationary solutions are always stable.

%%%%%%%%%%%%%%%%%%%%%%%%%%%%%%%%%%%
\section{Numerical method for model I}              %
%%%%%%%%%%%%%%%%%%%%%%%%%%%%%%%%%%%
\label{snumeric}

The two differential equations for growing and shrinking 
microtubules  in (\ref{pgl}) are of first order with respect  to the 
length  $l$ of  the microtubules
and first order in time.
 A straight forward  spatial discretization of such   first order
equations often leads  to numerical instabilities. Especially 
the equation for shrinking microtubules, cf. Eq.~(\ref{pgl2I}), has problematic stability properties.
For this reason we approximate the
solutions of Eqs.~(\ref{pgl}) by a two mode ansatz
\begin{mathletters}
\label{Pnl}
\begin{eqnarray}
p_{g}(l,t) &=&  \exp\left(-\frac{f_{cat}^{(0)}}{v_g} l \right) \left( \frac{\nu}{v_g} 
                 + F_{g}(l,t) \right ) \, , \\
p_{s}(l,t) &=&  \exp\left(-\frac{f_{cat}^{(0)}}{v_g} l \right) \left( \frac{\nu}{v_s} 
                 + F_{s}(l,t) \right ) \, ,
\end{eqnarray}
\end{mathletters}
where the first mode  describes just the stationary solution and the 
second one the oscillatory contribution.
This approximation becomes exact 
close to the threshold and this ansatz 
leads with Eqs.~(\ref{pgl}) to two differential equations for 
$F_{g,s}$  
\begin{mathletters}
\label{Fnl}
\begin{eqnarray}
\label{Fgnl}
\partial_t F_g &=& 
\left( f_{cat}^{(0)}-  f_{cat}\right ) \left( \frac{\nu}{v_g} + F_g \right) 
- v_g \partial_l  F_g \, , \\
\label{Fsnl}
\partial_t F_s &=&  f_{cat}\left( \frac{\nu}{v_g}+F_g \right )
  - \frac{v_s}{v_g} f_{cat}^{(0)} \left( \frac{\nu}{v_s}+  F_s \right )\nonumber \\
          && + v_s \partial_l  F_s \, .
\end{eqnarray}
\end{mathletters}
Both   fields may be expanded 
 with respect to the first two spatial  
Fourier modes $e^{i\,nlk}$ ($n=0,1$) 
\begin{mathletters}
\label{Flo}
\begin{eqnarray}
\label{Fglo}
F_g(l,t) &=& B(t) + \frac{1}{2}\left( C(t) e^{ikl}\, +\, C^{\ast}(t) e^{-ikl} \right)\, ,  \\ 
\label{Fslo}
F_s(l,t) &=& D(t) + \frac{1}{2}\left( H(t) e^{ikl}\,+\, H^{\ast}(t) e^{-ikl} \right) \, ,
\end{eqnarray}
\end{mathletters}
in order to remove the spatial dependence from  Eqs.~(\ref{Fnl}).
Herein the   wave number  is chosen at its
value at the threshold of the Hopf bifurcation, $k=\omega_c/v_g$. 
This ansatz  leads to  a set of ordinary differential equations for the
time dependent amplitudes  $B(t),\, C(t),\, D(t), \, H(t)$ that are described  in 
the following.

Due to Eq.~(\ref{boundpg}) one has the
 boundary condition $F_g(l=0,t)=0$  that gives  
the  relation  $ C_R = -B $, with $Re(C) = C_R$,
between these  two  functions.
Ansatz (\ref{Fglo}) together with equation (\ref{Fgnl})
leads to the relation 
\begin{equation}
Im(C) = C_I =  \frac{\nu}{k v_g^2 } \left( f_{cat} -f_{cat}^{(0)} \right ) \, ,
\end{equation}
and to the first order differential equation in $B$
\begin{equation}
\label{dB}
\partial_t B =  \left( f_{cat}^{(0)} -f_{cat} \right )  
                \left(  \frac{\nu}{v_g} +B \right) \, .
\end{equation}
Ansatz (\ref{Fslo}) in equation (\ref{Fsnl}) gives the set of coupled
differential equations 
\begin{mathletters}
\label{dHH}
\begin{eqnarray}
\partial_t D  &=& f_{cat}  \left(  \frac{\nu}{v_g} +B \right) 
                 -\frac{\nu f_{cat}^{(0)}}{v_g}
                -\frac{v_s} {v_g} f_{cat}^{(0)}\, D \, , \\
\partial_t H_R &=& -f_{cat}  \, B 
                 -\frac{v_s}{v_g} f_{cat}^{(0)} \, H_R
                - k v_s H_I  \, ,\\
\partial_t H_I &=&  f_{cat}  \, C_I
                 -\frac{v_s}{v_g} f_{cat}^{(0)} \, H_I
                + k v_s H_R  \, .
\end{eqnarray}
\end{mathletters}
With the periodic  $l$--dependence given in 
Eqs.~(\ref{Flo}) the integrals in Eq. (\ref{ctnl}) can 
be evaluated and one obtains the following 
differential equation for the tubulin--t dimer density
\begin{eqnarray}
\label{ctnum}
  \partial_t c_t  = - \gamma v_g  K(t)
                       -\alpha \gamma L(t) + \alpha c_{0c}\left(1+ \varepsilon \right )
                       - \alpha c_t \, ,
\end{eqnarray}
where  the coefficients are given by  
\begin{eqnarray}
  K(t) &=& \int_0^{\infty}\, dl\, p_g(l,t) =\frac{\nu}{v_g \delta} + \frac{k( k B 
   - \delta C_{I})}{\delta (\delta^2+k^2)} \, , \\
  L(t) &=& \int_0^{\infty}\,dl \,l\, \left[\, p_g(l,t)+p_s(l,t) \,\right ] \nonumber \\
              &=& \frac{\nu} {\delta^2}\left( \frac{1}{v_g}+ \frac{1}{v_s}\right )+
        \Delta \left( 3\,\delta^2\,k^2 B +k^4 B -2\,\delta^3\,k C_{I} \right. \nonumber \\
        &+&\left. \delta^2 (\delta^2-k^2) H_{R} -2\, \delta^3\,k H_{I}
               +(\delta^2+k^2)^2 D \right ) \, ,  
\end{eqnarray}
and where 
the abbreviations $\Delta= [\delta^2 (\delta^2+k^2)^2]^{-1}$ and 
 $\delta = \fcat/v_g$ have been introduced.
The reduced control parameter $\varepsilon=(c_0-c_{0c})/c_{0c}$
measures the difference between the  
tubulin dimer concentration  $c_0$ and the critical 
 one,  $c_{0c}$.
For $\varepsilon>0$ sustained
oscillations occur but they  are damped below threshold  $\varepsilon<0$.
For the numerical solution of model II we use either 
the same approximation scheme, where only the factors
of the  scheme take a different form, or in the absence of $p_s$ 
a direct spatial discretization provides also a stable algorithm.  
The five differential equations for model I in 
Eq.~(\ref{dB}),  Eq.~(\ref{dHH}) and  Eq.~(\ref{ctnum}) 
and the 
corresponding three differential equations for model II  
are integrated numerically
by  a second-order Runge-Kutta method
with a time step  $\Delta t =0.01$.

% ******************************************************************************
%   Figure 8
% ******************************************************************************
%
\begin{figure}
\epsfxsize 8.7cm \ifpreprintsty \epsfxsize 15.0cm \fi
\vspace{-3mm}
\epsfbox {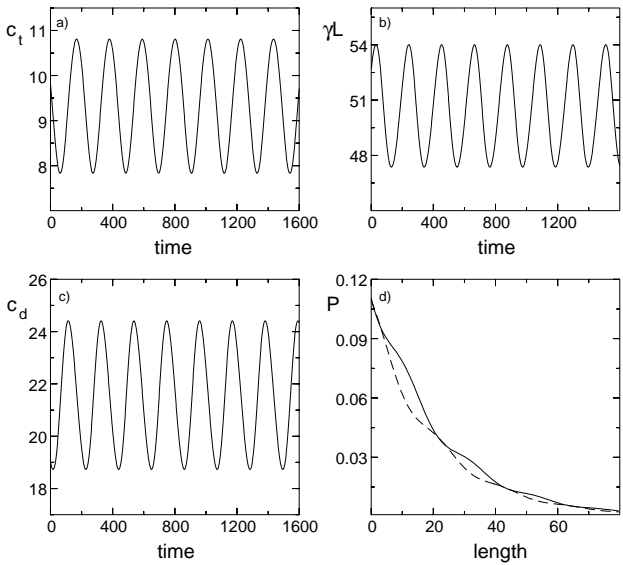}
\vspace{1mm}
\caption [Figure 8] 
  {For model I the time--dependence of the tubulin-t concentration 
 $c_t(t)$, the polymerized tubulin  $\gamma L(t)$
and the tubulin-d concentration $c_d(t)$ 
is shown in parts a), b) and c), respectively. The 
parameters  $\nu=0.01,\,\alpha=0.01, \, \beta=0.1$ were used
with the corresponding 
 critical initial concentration $c_{0c}=80.69$. 
The reduced control parameter is chosen at
the value $\varepsilon=0.01$.
 In part d) the length distribution of the
microtubule  $P(l)=p_g(l)+p_s(l)$  is shown at  two different times
 $t=876$ (solid) and $t=975$ (dashed), where
 $\gamma L(t)$ takes  its maximum and minimum,  respectively.
}
\label{fig8}
\end{figure}
%****************************************************************************

The time--dependence of the fields 
of  model I 
are shown in Figure \ref{fig8}  for one parameter set and  these fields 
obviously have  different phases.
The extrema (maxima)  of
the polymerized tubulin   $L(t)$ and the tubulin--d concentration 
$c_d(t)$ 
are delayed with respect to the extrema  of the tubulin--t concentration $c_t(t)$, a behavior
that is already indicated by  the reaction cycle shown 
in Fig.~\ref{figm1}.  At the  threshold the parameter dependence 
of these  phase differences may be calculated from the formulas 
given in  Sec.~\ref{LintravelI} and Sec.~\ref{Lintravell_II}.
In   Fig.~\ref{figphaseps} and in  Fig.~\ref{figphaseoli}
these phases are shown as function of 
the regeneration rate $\alpha$.
In Fig.~\ref{fig8}  the initial concentration
of tubulin  $c_0$ was chosen very close to the threshold
$c_{0c}$ with  
$\varepsilon=0.01$. At this value of the control parameter the 
oscillations behave harmonically  and the agreement 
between the numerical solution and the amplitude approximation
is rather good, as described in the following section.
For larger values of the reduced control parameter
the oscillations become  anharmonic.

% ******************************************************************************
%   Figure 9
% ******************************************************************************
%
\begin{figure}
\epsfxsize 8.7cm \ifpreprintsty \epsfxsize 15.0cm \fi
\vspace{-2mm}
\epsfbox {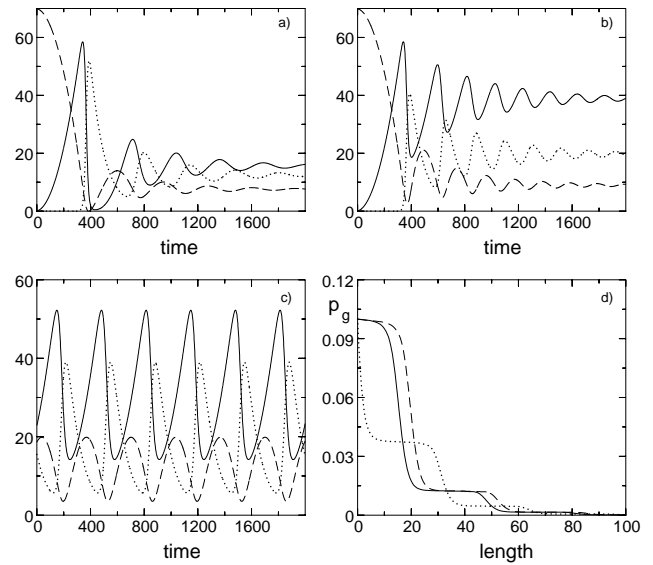}   
\vspace{1mm}
\caption [Figure 9] 
  {For model II the time--dependence 
of the polymerized tubulin $\eta \lambda L$ (solid),  $c_t$ (dashed)
  and $c_{oli}$ (dotted)  are shown 
below the Hopf bifurcation (see Fig.~\ref{fig3}) for $\alpha=0.0036$ in a) and
for $\alpha=0.08$ in b) with an initial concentration $c_0 =70$. The corresponding
critical concentration for both values of $\alpha$ is $c_{0c} = 134.15$.
 The stationary value of the tubulin-d concentration is
 $c_d^{(0)} = 2.465$ in a) and

$c_d^{(0)} = 34.305$ in b). 
In c) the time dependence is shown beyond 
the oscillation threshold at  $\alpha=0.01$.
In part d) the length distribution of the growing microtubules 
$p_g(l)$ is plotted at three different times. 
Both in c) and d)  the initial concentration is $c_0=80$ and the 
critical tubulin concentration  is $c_{0c}=65.97$ which
corresponds to the value $\varepsilon=0.21$ for the reduced control parameter.
The maximal length of the growing microtubules has been chosen as
$12 \cdot v_g/\fcat$.
In all parts the rest of the parameters are $\nu=0.01, \chi=0.01, v_g=0.1$. 
}
\label{fig9}
\end{figure}
%***********************************************************************************

As already been mentioned  in the introduction,
 the length distribution of 
filaments is a crucial difference between the biochemical reaction 
discussed in this work 
and the  common
oscillatory chemical reactions.
For model I we show in  Fig.~\ref{fig8}d) at two different
times  and at $\varepsilon=0.01$ 
the superposition of the length distribution  
of growing and shrinking microtubules, cf.   $P(l)=p_g(l)+p_s(l)$. 
The  exponential decay of the envelope of the length distribution
is described by 
Eq.~(\ref{eq:linpg}) and Eq.~(\ref{eq:linps}), with a decay rate $v_g/\fcat$,
and the modulation is due to the traveling waves 
in the time--dependent contribution.
The amplitude of the shrinking microtubules
is rather small for $\beta=0.1$, cf. Eq.~(\ref{eq:linps}), and 
therefore the contribution to $P(l)$ comes mainly from  growing
microtubules.

For model II   a discretization of the length coordinate in  the  equation 
for growing 
microtubule, cf. Eq.~(\ref{pgl1}),
  also provides a stable numerical algorithm.
Hence,  the nonlinear oscillatory solution 
can be obtained numerically  without the approximations 
as described for model I in Eqs.~(\ref{Pnl}) above.
The respective 
results are shown in Fig.~\ref{fig9}a)-c), where the  densities
$L, c_t$ and $c_{oli}$ are shown as function of time for three
different regeneration rates $\alpha$. For both  values of $\alpha$ in part a) and b) 
the tubulin concentration is smaller than the corresponding threshold value but
 the absolute distance $c_0 -c_{0c}$ 
to the threshold  is equal. These 
transient subthreshold--oscillations are remarkable, because 
the transient oscillations in experiments 
might  be subthreshold ones in contrast to the
common interpretation that the  oscillations are transient  
due to the   tubulin--t consumption during the microtubule polymerization.

  For the simulations shown in Fig.~\ref{fig9}
 a narrow length
 distribution $p_g$ has been used as initial condition. In
such cases
the tubulin-t concentration  corresponds almost to the total
initial concentration $c_{0}$. 
Starting with such an initial condition,  at first
tubulin--t dimers are consumed during the growth
of  microtubules. This leads to a first maximum  of
the polymerized tubulin  $L$, but 
the oligomers,  the decay product 
of the microtubules, are
negligible and as consequence the
densities of tubulin-d and tubulin-t drop down too.   But a small
$c_t$ increases 
the catastrophe rate $f_{cat}$ and
microtubules decay with a higher rate, which   increases the density of  oligomers etc..
 After a few of such oscillations the densities reach their
stationary values for subthreshold concentrations.
In the case of a large regeneration rate
the oscillation frequency is much higher 
than for small regeneration rates
and  more
oscillations are performed 
until the stationary values are reached.
The stationary value of the polymerized tubulin
 and of the oligomers is also much larger in part b) than in part a) whereas
the stationary value for the tubulin--t dimers remains nearly constant.
The reason is the low density  $c_d$  in the case of  large values of $\alpha$.

Far beyond the threshold of  the Hopf bifurcation, the oscillations
become anharmonic as shown in Fig.~\ref{fig9}c).
Hereby the oscillations of the total polymerization and of their
 decay product, $c_{oli}$, are nearly in antiphase, similar as  in experiments described in 
 Ref.\cite{Mandelkow:88.1}. At the threshold a phase difference of $\pi$
was only possible in the limit of large regeneration rates $\alpha$ and for a 
much smaller dissociation  rate $\chi$. If we consider
initial concentrations which are  much larger than the corresponding
critical concentration,
the phase shift of $\pi$ between $L(t) $ and $c_{oli}(t)$ is also
possible at intermediate values of $\chi$ and $\alpha$.

In  Fig.~\ref{fig9}d) the distribution 
for growing microtubules  $p_g(l)$  is shown for model II at three different
times. The reduced control parameter $\varepsilon=0.21$ 
is rather large  compared to its value in  Fig.~\ref{fig8} and 
the curves indicate that the length distribution of microtubules 
isn't described anymore by
harmonic  traveling waves as in the vicinity of the bifurcation.
As long as the tubulin-t density is large and the
  catastrophe rate  small,  microtubules grow with a
constant velocity $v_g$ and only a few of them
experience a catastrophe. During this period 
a plateau in the length distribution is build.
But after a large amount of tubulin-t has been used up
the catastrophe rate  $f_{cat}$ increases
very steeply leading to a 
strong decay of the microtubules at all lengths.
 If  $f_{cat}$ drops down again 
the low density of long microtubules grow further 
with a constant  velocity $v_g$ and 
short microtubules are nucleated at a higher density.
This leads to a  step in the distribution 
  that   travels with the growth 
velocity $v_g$ to larger values of $l$. Therefore,
far beyond the oscillatory 
threshold the temporally anharmonic behavior of $c_t$ 
leads in this manner to 
step--like length distribution of the microtubules.

%%%%%%%%%%%%%%%%%%%%%%%%%%%%%%
\section{Amplitude expansion}
\label{sampli}
%%%%%%%%%%%%%%%%%%%%%%%%%%%%%%

Here we focus  on a semi-analytical 
treatment of the  oscillating polymerization slightly beyond its onset. 
The interesting question here is whether
the bifurcation to these oscillations is
continuous (supercritical) or discontinuous (subcritical).
In physical systems Hopf bifurcations are mostly 
subcritical \cite{CrossHo,Zimmermann:93.1},  but for the models discussed in this work 
we   always find a supercritical one. This is  advantageous, because 
for a supercritical Hopf bifurcation 
a semi--analytical treatment is possible.
The appropriate frame work 
is the universal amplitude equation of oscillatory fields,
 cf.  Eq.~(\ref{ampliin}). This  equation will
 be derived  in the present  section 
from the basic  reaction equations of microtubule polymerization. 

The perturbation analysis employed for the  derivation of   Eq.~(\ref{ampliin})
is an expansion of the solutions of the basic equations with respect to 
small amplitudes of the oscillatory contributions
\cite{CrossHo,Strogatz:94}. As a small  
 perturbation parameter the relative difference 
between the actual  tubulin concentration $c_0$ and the 
critical tubulin concentration $c_{0c}$ is introduced
\begin{equation}
\label{reducedcontrol}
\varepsilon = \frac{c_0 - c_{0c}}{c_{0c}}\,.
\end{equation}
A signature  for  the $\pm$ symmetry of the oscillatory 
behavior  \cite{CrossHo,Strogatz:94} is the power law
for the oscillation amplitude
$A \sim \sqrt{\varepsilon}$.
Accordingly the solutions of the basic equations at the threshold  are
expanded   with respect to powers  of $\sqrt{\varepsilon}$
\begin{eqnarray}
{\bf u} = {\bf u}^{(0)} + \varepsilon^{1/2} {\bf u}^{(1)} + \varepsilon \, {\bf u}^{(2)} 
         + \varepsilon^{3/2} {\bf u}^{(3)} + {\cal O}(\varepsilon^2) \,\, ,
\end{eqnarray}
where  the 
vector notation ${\bf u}^{(j)} = (  \pgti{j}, \psti{j}, \ctti{j} )$ 
is used 
with  $j=0,1,2,3$. The components of ${\bf u}^{(j)}$
 differ by a factor of $\sqrt{\varepsilon}$, such as 
 $\sqrt{\varepsilon}\,  \ctti{1} = c_t^{(1)}$ etc. .  The components of ${\bf u}^{(0)}$ 
describe the  stationary microtubule polymerization as given 
 in Sec.~\ref{sstasol} and the components of 
 ${\bf u}^{(1)}$ describe  the  linear oscillatory 
solutions  that  may be written at the  threshold 
in the following form 
\begin{eqnarray}
\label{linamu1}
{\bf u}^{(1)}= B \, {\bf u}_{e} \,e^{i\omega_c t} + \it{c.c.} \,  .
\end{eqnarray}
Here  $\bf{u}_{e}$ includes 
the amplitude ratios between the fields  $c_t^{(1)}$,
$p_g^{(1)}$, $p_s^{(1)}$
at the threshold  and $\sqrt{\,\varepsilon} B = A$ as explained below.

Close to the threshold one has  $Re(\sigma) \sim \varepsilon  \ll 1$ and 
the linear solution ${\bf u}^{(1)} 
\sim e^{\sigma t}$ grows or decays  only by 
a very small amount   during  one  oscillation period $2\pi/ \omega_c$. 
These two disparate time scales near  the threshold,
that of the oscillation period ($\propto 2\pi/ \omega_c$) 
 and that of growth and decay ($ \propto 1/\varepsilon$), may be  separated  
within a 
perturbation expansion by introducing 
a slow time scale $T=\varepsilon t$
\cite{CrossHo,Strogatz:94}. The fast time scale 
is included in the exponential function $e^{i\omega_c t}$ 
and the other  one  will be described by a
slowly varying amplitude $B(T)$. Accordingly the linear solution
near threshold may be written as
\begin{eqnarray}
{\bf u}^{(1)}(t,T) = B(T)\, {\bf u}_{e} \,  e^{i\omega_c t} + \it{c.c.} \, .
\end{eqnarray}
In order to differentiate  this product of time dependent functions, instead of 
applying the chain rule of differentiation,  one may replace 
this operation by the following sum 
$\partial_t \to \partial_t + \varepsilon \, \partial_T$. Here
$\partial_t$ acts only  on the fast time--dependence
occurring in the  exponential function and $ \partial_T$
acts only  on the amplitude 
$B(T)$.

\begin{figure}
\epsfxsize 8.7cm \ifpreprintsty \epsfxsize 11.0cm \fi
\epsfig {file=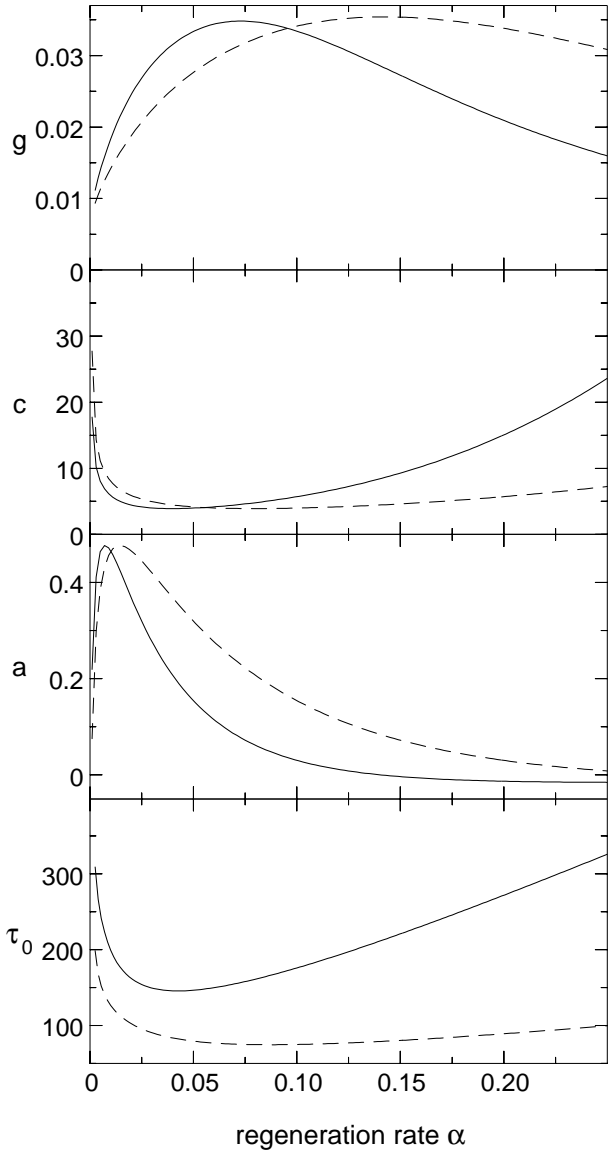,height=16cm}
\caption [Figure 5] 
  {The coefficients  $\tau_0$, $a$, $g$ and $c$ 
of the amplitude equation (\ref{amplps})  are shown for model I 
as function of the  regeneration rate $\alpha$
and for two different  nucleation rates 
$\nu=0.01$ (solid line) and  $\nu=0.04$ (dashed). For the rest of parameters the values
 $v_g=0.1$, $\beta=0.1$, $f=0.1$ and $c_{f}=3$ have been chosen.  
}
\label{fig5}
\end{figure}

\begin{figure}
\epsfxsize 8.7cm \ifpreprintsty \epsfxsize 11.0cm \fi
\epsfig {file=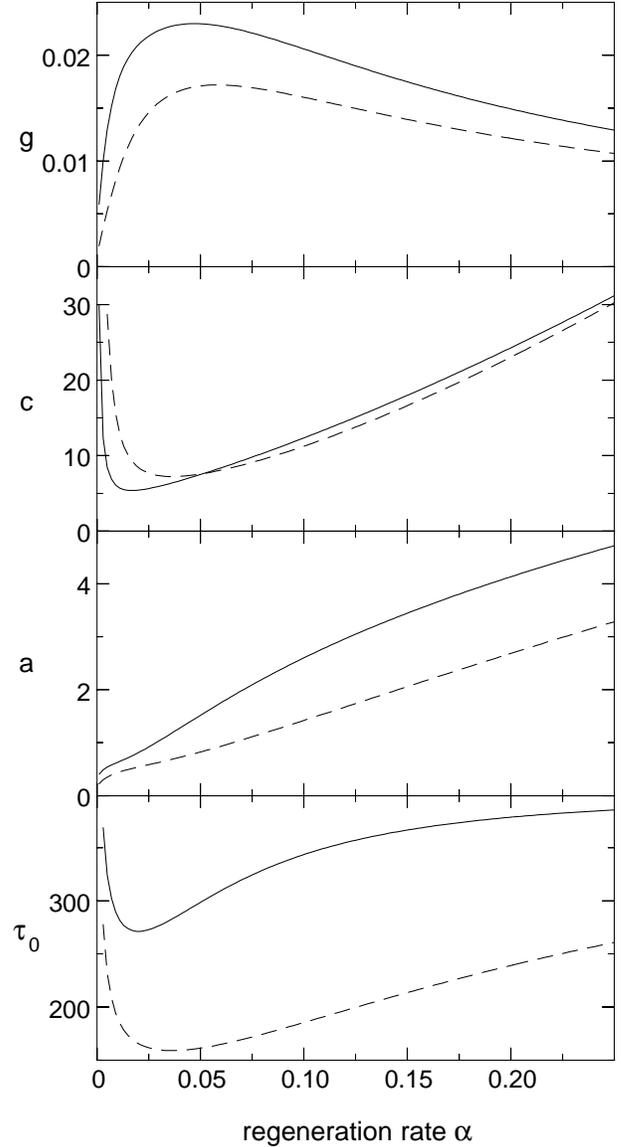,height=16cm}
\caption [Figure 7] 
  {
The linear and  nonlinear coefficients
of the amplitude equation  (\ref{amplps}) are shown for model II 
as a function of the   regeneration rate $\alpha$
and for two different nucleation rates   
$\nu=0.01$ (solid) and  $\nu=0.04$ (dashed).
For the rest of parameters the  values 
 $\chi=0.01$, $f=0.1$, $c_{f}=3$ have been chosen.  
}
\label{fig7}
\end{figure}

Using this replacement and the  $\varepsilon$-expansion of ${\bf u}$
the basic equations given in Sec.~\ref{smodel} 
can be ordered   with respect to  powers of $\sqrt{\varepsilon}$.
In this way  we obtain a hierarchy of partial differential equations,
which we need up to 
${\cal{O}}(\varepsilon^{3/2})$. The whole procedure  
is described in more detail in Appendix  \ref{appA}. 
The amplitude equation follows  from  a solubility condition 
for the equation at order  
${\cal{O}}(\varepsilon^{3/2})$ and it has the following form 
\begin{eqnarray}
\tau_0 \partial_{T} B = \left(1+ ia \right) B - g\left( 1 + ic\right) |B|^{2} B \,.
\end{eqnarray}
$\tau_0$ is the relaxation time, $a$ is the  linear and $c$ is the
nonlinear  frequency shift. The  nonlinear coefficient $g$
determines the bifurcation structure.  For $g>0$ the bifurcation 
is supercritical (steady) and for $g<0$ the bifurcation 
is subcritical (unsteady).
For the coefficients $\tau_0$, $a$, $g$ and $c$ one obtains 
long  expressions in terms of the reaction constants of the
basic equations, that have been calculated by 
 using computer algebra. The respective formulas are not presented but
the parameter dependence of the coefficients is shown 
in Fig.~\ref{fig5} for model I and in  Fig.~\ref{fig7} for model II.

Rescaling the time $T=\varepsilon t$ 
and amplitude $A =\sqrt{\,\varepsilon} \,B$
 back to the original units  
yields the amplitude equation  
\begin{eqnarray}
\label{amplps}
 \tau_{0}\partial_{t}A = \varepsilon \left ( 1 + ia \right ) A 
               - g \left(1 + i c\right) |A|^{2}A  \,\,, 
\end{eqnarray}
as introduced in Sec.~\ref{secintro}.  This equation has simple 
 nonlinear oscillatory solutions 
 of the form $A=F e^{i\Omega t}$, with an amplitude $F$ and
a frequency $\Omega$ as follows
\begin{eqnarray}
\label{ampfreq}
 F\,&=&\, \sqrt{\, \frac{\varepsilon}{g}} \quad , \quad \nonumber \\
         \Omega \,&=&\frac{1}{\tau_0} \left(\varepsilon a - gcF^2 \right)= 
\, \frac{a-c}{\tau_0} \, \varepsilon \, .
\end{eqnarray}
$ \Omega$ describes the deviation of the oscillation frequency  from the
critical one, $\omega_c$. 

The linear coefficients $\tau_0$ and $a$ of Eq.~(\ref{amplps}) 
may directly be calculated from the dispersion relation in 
Eq.~(\ref{dispers1})  or  Eq.~(\ref{disoli2}) in the following way. 
The solution $A=0$ of Eq.~(\ref{amplps}) corresponds to
the stationary  polymerization described in Sec.~\ref{sstasol}, which  
is  in the range $\varepsilon <0$ stable
 against small perturbations
 $ A \sim \tilde F e^{\sigma t} $ ( with $ \tilde F \ll \sqrt{|\varepsilon|}$)
and unstable for $\varepsilon >0$. 
Neglecting in   Eq.~(\ref{amplps})   the contributions due to the cubic nonlinearity 
one obtains from  its  linear part  
 the dispersion  relation  $\sigma = \varepsilon (1 +i a) / \tau_0$. This formula gives 
the relaxation time $\tau_0$ and the linear frequency dispersion $a$ 
in terms of derivatives of the growth rate with respect to the control parameter $\varepsilon$: 
  $ \tau_0 = ( \partial Re(\sigma)/\partial \varepsilon )^{-1}$
and $ a = \tau_0  \partial Re(\sigma)/ \partial  \varepsilon$.
If  $\varepsilon$ is expressed in terms of  the dimer density 
$c_0$, cf. Eq.~(\ref{reducedcontrol}),  then  both quantities may also  be
written in terms of the derivatives with respect to $c_0$ 
\begin{eqnarray}
\label{folinkof}
 \tau_0 \,=\, \frac{1}{c_{0c} \partial Re(\sigma) / \partial c_0}
              \quad , \quad a\,=\, c_{0c} \tau_0 \frac{\partial Im(\sigma)}
                 {\partial c_0} \,\, ,
\end{eqnarray}
whereby both derivatives are taken at the  threshold concentration $c_{0c}$.
The  dispersion relation $\sigma(\varepsilon)$ 
as  obtained on the one hand by the  amplitude equation 
and on the other hand by solving   Eq.~(\ref{dispers1}) 
or Eq.~(\ref{disoli2}), both have to reproduce 
the growth or decay dynamics of small perturbations with respect to
the stationary polymerization. 
Therefore, the coefficients $\tau_0$ and $a$ of the amplitude equation 
can directly be calculated 
via Eq.~(\ref{folinkof}) from the numerical solutions 
of  Eq.~(\ref{dispers1}) 
or Eq.~(\ref{disoli2}).

One aim of the 
amplitude expansion is the determination of the type
of the Hopf bifurcation.   
For two different nucleation rates $\nu=0.01$ and $\nu=0.04$
the variation of the nonlinear coefficient 
 $g$ as function of the regeneration rate $\alpha$ 
 is shown for model I  in  Fig.~\ref{fig5} (top)  and for model II with
oligomer dynamics  in
 Fig.~\ref{fig7}. In both cases $g$ behaves  rather similar and
$g$ is  positive for the models  investigated 
in this work. Therefore the Hopf bifurcation is  supercritical. 
In   Fig.~\ref{fig5} and Fig.~\ref{fig7} the nonlinear coefficient $g$
increases at  first with the regeneration rate $\alpha$ and reaches a
maximum in a range  where the threshold concentration 
$c_{0c}(\alpha)$ takes its minimum. 
The corresponding threshold curves $c_{0c}(\alpha)$ for two different
nucleation rates $\nu=0.01$ and $\nu=0.04$ also cross each other. 
Beyond this maximum of $g$ ( where
$c_{0c}(\alpha)$ takes its minimum)  decreases again.

It should be mentioned  that
for a given value of the control parameter  $\varepsilon$
a large value of $g$ corresponds to a small value of the oscillation
amplitude. 
Since the threshold $c_{0c}(\alpha)$ varies too, 
the variation of the oscillation amplitude with $\alpha$  is much less 
when $c_{0c} \varepsilon$ is kept constant. 
According to the sign of the nonlinear coefficient $c$ the
oscillation frequency $\omega_c + \Omega$ decreases with increasing
values of $\varepsilon$. The results in terms of the amplitude 
equation are in fairly good agreement with the
behavior of the full numerical solution of the
basic reaction equations. 

% ******************************************************************************
%   Figure 10
% ******************************************************************************
%
%****************************************************************************

A  determination of the bifurcation structure
by numerical simulations  of the basic equations is 
error prone compared to results of perturbation calculation 
described here. Besides the lower accuracy, parameter studies
such as in  Fig.~\ref{fig5} and Fig.~\ref{fig7} 
are with numerical simulations   much more time consumptive.

% ******************************************************************************
%   Figure 11
% ******************************************************************************
%
%**************************************************************************** 

\begin{figure} %[b]
\epsfxsize 8.7cm \ifpreprintsty \epsfxsize 15.0cm \fi
\epsfbox {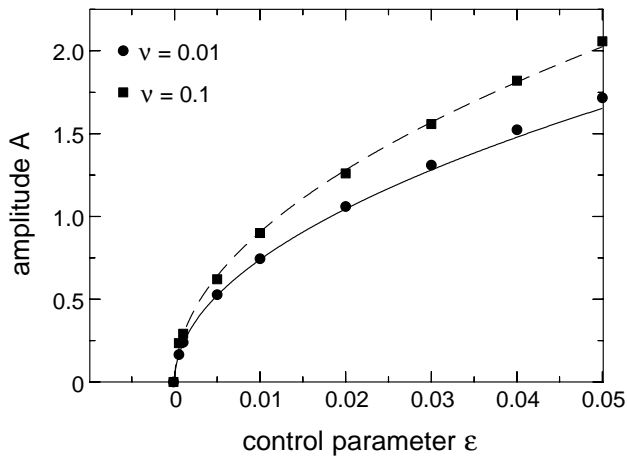}
\caption [Figure 3] 
  {The amplitude of the oscillations as function
of the reduced control parameter $\varepsilon$ and for two different
nucleation rates $\nu$. The solid line 
is the result of the amplitude equation and
data points have been determined from simulations 
as described in section \ref{snumeric}.  
The parameters that have been used 
are $\gamma=1$, $\alpha=0.01$, $v_g=0.1$, $\beta=0.1$, $f=0.1$, $c_{f}=3$.
}
\label{fig6}
\end{figure}

Close to threshold the advantages of the perturbation calculation
are obvious. 
However, it is a priori unknown 
in which  $\varepsilon$-range the amplitude equation approach  (\ref{amplps}) applies quantitatively. 
For some systems the  
amplitude equation is valid in a rather large range of the control
parameter  $\varepsilon$, but 
for other systems its validity is restricted to very small values of it, cf. Ref.~\cite{CrossHo}.
In order to check this for  our models of microtubule polymerization, 
we  compare in  
 Fig.~\ref{fig6} 
the variation  of the oscillation amplitude of
 $c_t^{(1)}$ with the control  parameter  $\varepsilon$ as obtained by
 the numerical solution  described in Sec.~\ref{snumeric} 
and by the 
solution $A = \sqrt{\varepsilon/g\,}$ 
of the amplitude equation for two different values of the
nucleation rate $\nu$. At larger values of the control parameter,  $\varepsilon=0.1$,
 the  difference between the results for the numerical solution and the 
amplitude equation is still less than $8 \%$. For model II the deviations
are larger between the amplitude determined by the amplitude equations approach 
and the numerical solutions with ansatz in Eq.~(\ref{Flo}). However, when
we solve the equation for growing microtubule, cf. Eq.~(\ref{pgl1}),
 numerically by discretization of the length coordinate, 
the deviations becomes smaller.

% ******************************************************************************
%   Figure 12
% ******************************************************************************
%
%*********************************************************************

For both  models, the linear coefficient $\tau_0$
and the nonlinear coefficients $g, c$ don't differ 
very much from each other. However, the linear frequency shift $a$ increases
in model II for large regeneration rates $\alpha$ whereas for model I
it  decreases.  Nevertheless  in both models the nonlinear 
frequency correction due to the values of $c$ is much larger 
than the linear correction due to 
$a$, $c>>a$.

Whether the bifurcation to oscillatory  polymerization is 
also supercritical in experiments under 
stationary regeneration conditions,  is an open question. 
Therefore an  experimental determination of the bifurcation--type 
 would be  an important test for the reduced models
investigated in this work.

The variation of the other coefficients with the regenerate $\alpha$ 
provides further contact between the model parameters and
experimentally measurable quantities. 
The parameters $c$, $a$ and $\tau_0$ may be
determined as follows. Studying the growth of small perturbations, 
$c_t^{(1)} \propto e^{Re(\sigma) t} =e^{\varepsilon t /\tau_0 }$ by plotting the logarithm 
$\varepsilon t /\tau_0 \propto  \log( c_t^{(1)})$ as function of 
time and for different values
of $\varepsilon$, the relaxation time $\tau_0$ may be determined. 
In a similar  manner $a$ may be determined by  studying the frequency
of a perturbation far below its nonlinear  saturation amplitude.
If the perturbation saturates finally, the oscillation frequency of the 
nonlinear solution 
changes with $\varepsilon$ as indicated by Eq.~(\ref{ampfreq}). From this 
$\varepsilon$--dependence  the nonlinear coefficient $c$ may be extracted.
An experimental determination of these coefficients, as described,  would 
be a further test of the basic model equations.

%%%%%%%%%%%%%%%%%%%%%%%%%%%
\section{Summary and Conclusion}        %
\label{summary}
%%%%%%%%%%%%%%%%%%%%%%%%%%%

Two reduced models, that capture oscillating microtubule polymerization
and the length distribution of the microtubule filaments,  
as described in Sec.~\ref{smodel}, 
have been analyzed. 
In both  models  the complex  biochemical reaction steps 
 of microtubule polymerization are
described by a few ones, which have been identified in 
experiments to be  important. 
The  focus on a few  essential degrees of 
freedom leads to some simplicity of the models 
that allows for instance a derivation of analytical expressions for the
 threshold
and the oscillation frequency at the Hopf bifurcation.
Such  analytical results make  trends as function of the
reaction rates more easily visible.  Some of these trends  may  be tested in experiments
and  some of the reaction constants
may be measured.

At threshold also analytical expressions could be derived
for the temporal evolution of the 
  concentrations and the length distribution of microtubules.
 Those  
provide a detailed  picture about the temporal variation of the fields, their relative phases and
the amplitude ratios  between them. The formula for the  
length distribution is especially instructive, cf. Eq.~(\ref{eq:linpg}),
it   describes a superposition of  amplitude oscillations of the distribution  
and  traveling waves,  where  the waves always travel towards 
larger lengths. This qualitative behavior of the length distribution
during oscillatory polymerization  
is rather independent of the respective model and it is a rather
general feature. 
 A few snap shots  of the numerically
generated  distribution  far beyond threshold are 
 shown in Fig.~\ref{fig9}b). The distribution 
includes still traveling waves, but far beyond threshold these
behave rather anharmonic.

For stationary reaction conditions,  as 
assumed in this work, Fig.~\ref{fig9} shows a remarkable result.
In part a) and b) of this figure a subthreshold concentration for tubulin 
was assumed, i.e. $c_0 < c_{0c}$, and   in both cases  the final state is 
 stationary polymerization. However,  on the route to the stationary state
transient oscillations occur.  Therefore,  the transient character of the microtubule oscillations 
 observed in an experiment  with an enzymatic regeneration process 
for GTP  \cite{Job:89.1}, could have, according to the results described in this work, 
 its origin in a to low tubulin concentration. A higher tubulin concentration or
an appropriate  regeneration rate of GTP and a different live time of oligomers 
could lead in a similar experiment to persistent microtubule oscillations. 

For transient oscillations observed in other experiments  the 
common interpretation is as follows.
During the microtubule polymerization in experiments  the available 
 GTP  is  used up and the oscillations  last only
for a few periods.  If GTP is continuously 
supplied, the simultaneously increasing amount of GTD inhibits  
various reactions steps and slows down the reaction cycle. The results shown 
in  Fig.~\ref{fig9} a) and   Fig.~\ref{fig9} b) indicate that 
oscillations may occur as a transient because either
GTP is used up or  the tubulin concentration has only a subthreshold value.
 Accordingly,  there may be several 
 reasons for transient oscillations 
in experiments. Either the initial tubulin concentration has 
a subthreshold value, 
 the decay rate of oligomers and the regeneration rate
for GTP  do not have their optimal values 
 which would explain transient oscillations 
in experiments with a regenerative enzyme system, 
the reaction conditions are not constant, because 
 GTP is used up. 
 In the latter   case the
available tubulin--t decreases with time and the microtubule polymerization decays.

In an in vitro  experiment with constant reaction conditions the 
threshold of  oscillations might be measured by
increasing  the tubulin dimer concentration by appropriate steps. 
Immediately after each step transient oscillations might occur,
but  the threshold is only crossed  when the oscillations persist over a long time. 

In order to avoid numerical instabilities during long time simulations 
of the reaction equations, including Eq.~(\ref{pgl2I}), we 
use analytical approximations for the length 
dependence of the microtubule distributions as described in 
Sec.~\ref{snumeric}.  This stable numerical scheme
can  be generalized in future work to an effective algorithm 
for dealing with microtubule polymerization in 
   one and two spatial dimensions \cite{Hammele:02.120}
in  order to investigate spatial patterns occurring during polymerization of microtubules 
\cite{Mandelkow:89.1,Tabony:90.1}. 
 The respective extension of the amplitude equation
may also  lead to new interesting insights.

Microtubule filaments at a high  density show   a isotropic--nematic phase transition 
\cite{Hitt:90.1}, similar as  it has been observed for   F-actin filaments 
 \cite{Suzuki:91.1}.
Since the  early theory of Onsager \cite{Onsager:49.1} 
this transition for rods in a solvent is a well understood  phenomenon 
\cite{deGennes:93}. 
For a monodisperse distribution of 
filaments of fixed length, i.e. without polymerization kinetics, 
many aspects of the isotropic--nematic transition have been understood. 
For polydisperse rod mixtures some aspects of the isotropic-nematic 
 transition can be considered to be understood too \cite{Vroege:92.1}.
 The effects of nucleation, growth of filaments and  the decay of filaments 
on the isotropic--nematic transition are not known at present 
and one may expect interesting phenomena related to this kinetics
\cite{Ziebert:02.120}. 
The effect of oscillating microtubule polymerization on the
isotropic--nematic transition is also completely  unexplored 
at present and will be investigated in forthcoming works
\cite{Ziebert:02.120}.

%amplitude equations supercritical ...

It is a great pleasure to thank M. Breidenich, H. Flyvbjerg, E. Mandelkow, H. M{\"u}ller--Krumbhaar,
J. Prost and J. Tabony for interesting  discussions.

%
%*******************************************************************
%   Appendixes
%*******************************************************************
%
\appendix
\section{Amplitude expansion for model I}
\label{appA}
For model I and  the catastrophe rate given in Eq.~(\ref{rateexp}) 
the major steps of the derivation of the amplitude equation (\ref{ampliin}) 
are described in this appendix. Since we neglect the
rescue of shrinking microtubules, $f_{resc}=0$, 
the only nonlinear term in the basic equations of model I is the product
$f_{cat}(c_t) \, p_g$ in Eq.~(\ref{pgl}). 
At first we expand the concentrations $c_t, c_0$ and length
distributions $p_{g,s}$ with respect to deviations from their stationary 
value at the threshold for oscillation, such as for instance 
for  the tubulin-t concentration, i.e.  
 $c_t-c_t^{(0)}= \sqrt{\varepsilon} \ctti{1}  + \varepsilon \ctti{2} + \ldots $. 
Note that the tilded fields differ just by
a power of  $\sqrt{\varepsilon} $  from the untilded
fields as introduced  in Sec.~\ref{sthreshosz}, cf.  $( \sqrt{\varepsilon})^j  \ctti{j}=c_t^{(j)}$ etc.  .
In order to simplify the notation of this appendix we drop the ``tilde'' 
 and the expansion of the catastrophe rate
takes the form

\begin{eqnarray}
f_{cat} = f_{cat}^{(0)} + \varepsilon^{1/2}f_{cat}^{(1)} 
   +\varepsilon f_{cat}^{(2)} + \varepsilon^{3/2}f_{cat}^{(3)} + \ldots\,\, ,
\end{eqnarray}
whereby the coefficients of this  expansion are
\begin{mathletters}
\begin{eqnarray}
f_{cat}^{(1)} &=& - f_{cat}^{(0)} \frac{\ct{1}}{c_f} \, ,\\
f_{cat}^{(2)} &=& f_{cat}^{(0)} \left( \frac{1}{2} \left( \frac{\ct{1}}{c_f}\right)^2
               -\frac{\ct{2}}{c_f} \right) \, ,\\
f_{cat}^{(3)} &=& f_{cat}^{(0)} \left( \frac{\ct{1}\ct{2}}{c_f^2}
                - \frac{\ct{3}}{c_f}- \frac{1}{6} 
                \left( \frac{\ct{1}}{c_f}\right)^3 \right) \, .
\end{eqnarray}
\end{mathletters}

Collecting in  Eq.~(\ref{pgl}) 
and Eq.~(\ref{ceqsh}) the contributions to the order  $\varepsilon^{1/2}$ we recover
the linear equations given in Sec.~\ref{sthreshosz} 
\begin{mathletters}
\label{amplpg1}
\begin{eqnarray}
  \label{pg1}
  \partial_t p_g^{(1)} &=& f_{cat}^{(0)} \frac{\ct{1}}{c_f}p_{g}^{(0)}
                         -f_{cat}^{(0)} \, p_g^{(1)} - v_g\partial_l p_g^{(1)}\, ,  \\
  \label{ps1}
  \partial_t p_{s}^{(1)} &=& - f_{cat}^{(0)} \frac{\ct{1}}{c_f}p_{g}^{(0)} 
                           + f_{cat}^{(0)} \, p_g^{(1)}   
                          + v_s\partial_l p_{s}^{(1)} \, ,\\
   \label{ct1}
  \partial_t c_{t}^{(1)} &=&  -\gamma \int_0^{\infty} dl \left( v_g p_g^{(1)} 
                         + \alpha l (p_g^{(1)}+p_{s}^{(1)}) \right ) \nonumber \\
                     && \hspace{1cm}    - \alpha \ct{1} \, .
\end{eqnarray}
\end{mathletters}
At order  $\varepsilon$ we obtain the three equations 
\begin{mathletters}
\begin{eqnarray}
  \label{pg2}
\partial_t p_{g}^{(2)} &=& \left(  f_{cat}^{(0)} \frac{\ct{2}}{c_f}p_{g}^{(0)}
- f_{cat}^{(0)} \, p_{g}^{(2)} - v_g\partial_l p_{g}^{(2)} \right ) \nonumber \\
                    &&    + f_{cat}^{(0)} \frac{\ct{1}}{c_f} p_g^{(1)} 
                        - \frac{f_{cat}^{(0)}}{2}\left(\frac{\ct{1}}{c_f} \right )^2
                           p_{g}^{(0)}  \, ,\\
 \label{ps2}
\partial_t p_{s}^{(2)} &=& \left(- f_{cat}^{(0)} \frac{\ct{2}}{c_f}p_{g}^{(0)}
              + f_{cat}^{(0)} \, p_{g}^{(2)} + v_s\partial_l p_{s}^{(2)} \right )\nonumber \\  
                    &&  - f_{cat}^{(0)} \frac{\ct{1}}{c_f} p_g^{(1)} 
                        + \frac{f_{cat}^{(0)}}{2}\left(\frac{\ct{1}}{c_f} \right )^2
                          p_{g}^{(0)} \, , \\
\label{ct2}
\partial_t \ct{2} &=&  - \gamma \int_0^{\infty} dl \left( v_g p_{g}^{(2)} 
                       + \alpha l (p_{g}^{(2)}+p_{s}^{(2)}) \right ) \nonumber \\
                     &&  + \alpha c_{0c} - \alpha \ct{2} \, .
\end{eqnarray}
\end{mathletters}
With the solutions of the equations at the previous 
order  ${\varepsilon}^{1/2}$,  that are already given in
Sec.~\ref{LintravelI},  the equations at order $\varepsilon$ 
have to be solved. These solutions have the following form 
\begin{mathletters}
\begin{eqnarray}
  \ct{2} &=& A_0 + A_2 \exp{(2i\omega_c t)} + c.c.\,,\\
  p_g^{(2)} &=& e^{-{f_{cat}^{(0)}l}/{v_g}} \left( B_0(l) + 
\left[ B_2(l) e^{2i\omega_c t} + c.c.\right] \right )\,, \\
  p_s^{(2)} &=&   e^{-{f_{cat}^{(0)}l}/{v_g}} \left( D_0(l) + 
\left[ D_2(l) e^{2i\omega_c t} + c.c. \right] \right ) \,,
\end{eqnarray}
\end{mathletters}
whereby the expressions for the coefficients $A_0$, $A_2$, $B_i$ and $D_i$ are rather 
lengthy in terms of the 
coefficients of the solutions at order  ${\varepsilon}^{1/2}$ 
and are not given here.

The equations at the next higher order  ${\varepsilon}^{3/2}$ are 
\begin{mathletters}
\begin{eqnarray}
  \label{pg3}
\partial_T p_g^{(1)}+ \partial_t p_{g}^{(3)} &=& \left(  f_{cat}^{(0)} 
               \frac{\ct{3}}{c_f}p_{g}^{(0)} - f_{cat}^{(0)} \, p_{g}^{(3)} 
                - v_g\partial_l p_{g}^{(3)} \right )\nonumber \\
          && - \left( f_{cat}^{(0)} \frac{\ct{1}\ct{2}}{c_f^2} -\frac{f_{cat}^{(0)}}{6}
                 \left(\frac{\ct{1}}{c_f}\right)^3 \right ) p_{g}^{(0)} \nonumber \\
          && + \left( f_{cat}^{(0)} \frac{\ct{2}}{c_f} - \frac{f_{cat}^{(0)}}{2}
              \left( \frac{\ct{1}}{c_f} \right )^2 \right ) p_g^{(1)}\nonumber \\
          &&             + f_{cat}^{(0)} \frac{\ct{1}}{c_f} p_{g}^{(2)} \, ,  \\
\label{ps3}
\partial_T p_{s}^{(1)}+ \partial_t p_{s}^{(3)} &=& \left(-f_{cat}^{(0)} 
               \frac{\ct{3}}{c_f} p_{g}^{(0)} + f_{cat}^{(0)} \, p_{g}^{(3)} 
                + v_s \partial_l p_{s}^{(3)} \right ) \nonumber \\
         && + \left( f_{cat}^{(0)} \frac{\ct{1}\ct{2}}{c_f^2} -\frac{f_{cat}^{(0)}}{6}
                  \left(\frac{\ct{1}}{c_f}\right)^3 \right ) p_{g}^{(0)} \nonumber \\
         && - \left( f_{cat}^{(0)} \frac{\ct{2}}{c_f} - \frac{f_{cat}^{(0)}}{2}
               \left( \frac{\ct{1}}{c_f} \right )^2 \right ) p_g^{(1)} \nonumber \\
         &&             - f_{cat}^{(0)} \frac{\ct{1}}{c_f} p_{g}^{(2)} \, , \\
\label{ct3}
\partial_T \ct{1} + \partial_t \ct{3} &=&  - \gamma \int_0^{\infty} dl 
             \left( v_g p_{g}^{(3)} + \alpha l (p_{g}^{(3)}+p_{s}^{(3)}) \right )\nonumber \\
         &&   - \alpha \ct{3} \, .
\end{eqnarray}
\end{mathletters}
The two fields   $p_{g}^{(3)}$ and  $p_{s}^{(3)}$ have to be calculated 
explicitly at this order  from 
Eq.~(\ref{pg3}) and Eq.~(\ref{ps3}) as well. With both solutions  
the integral on the right hand side of Eq.~(\ref{ct3}) can be calculated. 
Eq.~(\ref{pg3}) and Eq.~(\ref{ps3}) include both contributions proportional to
$e^{i\omega_c t}$ and $e^{3 i\omega_c t}$, but only 
the single harmonic terms are relevant in Eq.~(\ref{ct3}). 
The coefficient 
of $e^{i\omega_c t}$ in Eq.~(\ref{ct3})  must vanish. Part of it vanishes automatically, because
it reproduces the threshold condition and the rest provides the amplitude equation 
with all the coefficients now given in terms of the reaction rates of the 
basic equations.

%
% ******************************************************************
%   References
% ******************************************************************
%
\bibliographystyle {prsty}

\end{multicols} 
\end{document}